\documentclass[twocolumn]{aa}
\usepackage{graphicx}
\usepackage{psfig}
\def\sec{\ifmmode {}^{\prime\prime}\else ${}^{\prime\prime}$\fi~}

\def\magdot{\ifmmode {}^{\rm m}\!\!\!.\, \else ${}^{\rm m}\!\!\!.\,$\fi}
\def\daydot{\ifmmode {}^{\rm d}\!\!\!.\, \else ${}^{\rm d}\!\!\!.\,$\fi}
\def\asec{\ifmmode ^{\prime\prime}\else$^{\prime\prime}$\fi}

\begin{document}
   \title{Integral field spectroscopy of the ultraluminous X-ray source Holmberg II X-1}

   \author{I. Lehmann
          \inst{1}
          \and
          T. Becker\inst{2}
          \and 
          S. Fabrika\inst{3}
          \and
          M. Roth \inst{2}
          \and
          T. Miyaji \inst{4}
          \and 
          V. Afanasiev \inst{3}
          \and
          O. Sholukhova\inst{3}
          \and
          S. F. S\'anchez  \inst{2}
          \and
          J. Greiner\inst{1}
          \and
          G. Hasinger\inst{1}
          \and
          E. Costantini\inst{1}
          \and
          A. Surkov\inst{3}
          \and
          A. Burenkov \inst{3,}\inst{5}
          }

   \offprints{I. Lehmann}

   \institute{Max-Planck-Institut f\"ur extraterrestrische Physik,
              Giessenbachstra\ss e, PF 1312, D-85741 Garching, Germany\\
              \email{ile@mpe.mpg.de, jcg@mpe.mpg.de, grh@mpe.mpg.de, elisa@mpe.mpg.de}
             \and
             Astrophysikalisches Institut Potsdam,
             An der Sternwarte 16, D-14471 Potsdam, Germany \\
             \email{tbecker@aip.de, mmroth@aip.de, ssanchez@aip.de}
             \and
             Special Astrophysical Observatory of the Russian AS, Nizhnij Arkhyz 369167, Russia\\
             \email{fabrika@sao.ru, olga@sao.ru, vafan@sao.ru, sura@sao.ru, ban@sao.ru}
             \and 
             Department of Physics, Carnegie Mellon University,
             Pittsburgh, PA 15213, USA\\
             \email{miyaji@astro.phys.cmu.edu}
             \and
             Isaac Newton Institute of Chile, SAO Branch, Russia\\
             }
   \date{Received -; accepted -}

   \abstract{ We present optical integral field observations of the H II region containing the ultraluminous X-ray source Holmberg II X-1. We confirm the existence of an X-ray ionized nebula as the counterpart of the source due to the detection of an extended He II $\lambda4686$ region (21$\times$47 pc) at the {\it Chandra} ACIS-S position. 
An extended blue objects with a size of 11$\times$14 pc is coincident with the X-ray/He II $\lambda4686$ region,
which could indicate either a young stellar complex or a cluster. 
We have derived an X-ray to optical luminosity ratio of L$_{X}/$L$_{B}\ge170$, and presumable it is L$_{X}/$L$_{B}\sim300-400$ using the recent HST ACS data.
We find a complex velocity dispersion at the position of the ULX. In addition, there is a radial velocity variation in the X-ray ionized region found in the He II emission of $\pm50$ km s$^{-1}$ on spatial scales of 2-3$^{\prime\prime}$. We believe  that the putative black hole not only ionizes the surrounding HII gas, but also perturbs it dynamically (via jets or the accretion disk wind). The spatial analysis of the public {\it Chandra} ACIS-S data reveals a point-like X-ray source and marginal indication of an extended component ($\ll 15$ \% of the total flux). The {\it XMM-Newton} EPIC-PN spectrum of HoII X-1 is best fitted with an absorbed power-law in addition to either a thermal thick plasma or a thermal thin plasma or a multi-colour disk black body (MCD). In all cases, the thermal component shows relatively low temperatures (kT$\sim0.14-0.22$ keV). Finally we discuss the optical/X-ray properties of HoII X-1 with regards to the possible nature of the source. The existence of an X-ray ionized nebula coincident with the ULX and the soft X-ray component with a cool accretion disk favours the interpretation of an intermediate-mass black hole (IMBH). However the complex velocity
behaviour at the position of the ULX indicate a dynamical influence of the black hole onto the local HII gas.
}
\titlerunning{Integral field spectroscopy of HoII X-1}
   \maketitle
%

   \begin{figure}[t]
   \centering
   \psfig{file=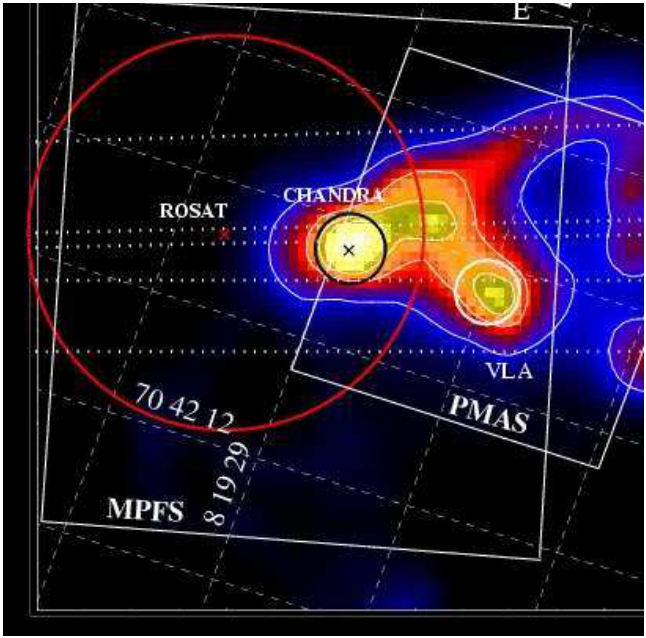,bbllx=40pt,bblly=30pt,bburx=482pt,bbury=470pt,width=8.5cm,clip=}       
      \caption{CFHT archival $H_{\alpha}$ image of the HoII X-1 region. The MPFS FOV and the PMAS mosaic FOV are superposed. The small black and the large red circles give the 90 \% confidence circles of the Chandra ACIS-S (RA: 8$^{h}$ 19$^{m}$ 29.0$^{s}$, DEC: +70$^{\circ}$ 42$^{\prime}$ 19.0$^{\prime\prime}$; J2000) and the ROSAT HRI positions, respectively. The small white circle shows the location of the radio peak at 6, 20, and 90 cm wavelengths (see Tongue \& Westpfahl \cite{Ton95}). Long-slit positions (1-5) are indicated with dotted lines. The overlayed coordinate grid is given in steps of 1$^{s}$ in RA and in steps of 4$^{\prime\prime}$ in DEC.}
         \label{FOV}
   \end{figure}

\section{Introduction}

The two main classes of discrete galactic X-ray sources, X-ray binaries and supernova remnants (SNR),
have been known and relatively well understood for decades. The third class of  galactic X-ray sources was detected with Einstein due to its high spatial resolution and its large collecting area (Fabbiano \cite{Fab89}). 
These objects have become known as Ultraluminous X-ray sources (ULX) and they have 0.5-10 keV luminosities of 10$^{39-41}$ erg s$^{-1}$, generally higher than black hole binaries such as Cyg X-1 and SMC X-1, but lower than that of active galactic nuclei (AGN). Assuming Eddington luminosities, this corresponds to accretion onto black holes
of masses between ten and several hundred solar masses suggesting intermediate-mass black holes (IMBHs; see Colbert \& Mushotzky \cite{Col99}, Miller \& Colbert \cite{Mil04} for a review). ULX are not located in the dynamical center of their host galaxies and thus they are not caused by sub-Eddington accretion onto a central AGN-type super-massive black hole.  
The identification of the optical counterparts of ULX is one of the most
important issues to determine the nature of these objects. The number of optically identified ULX is still
limited, only a small number of reliable optical counterparts are known to date. 
Several ULX seem to
be associated with H II regions or nebula (e.g., Pakull \& Mironi \cite{Pak01}, Foschini et al. \cite{Fos02}, and
Wang \cite{Wa02}).
However, some ULX have been associated with an accreting black hole in a globular cluster (see Angelini et al. \cite{Ang01}, Wu et al. \cite{Wu02}). The optical counterpart to an extremely luminous X-ray source near Holmberg IX is a shock-heated nebula, which is associated with a optically faint non-stellar source (Miller \cite{Mil95}).

The discovery of the intense He II $\lambda4686$ nebular recombination line in Holmberg II X-1 (hereafter; HoII X-1) indicates that
the interstellar medium probably reprocesses part of the X-ray luminosity of $\sim 10^{40}$ erg s$^{-1}$.  Assuming quasi-isotropic emission (Pakull \& Mironi \cite{Pak01}) and the distance of $D=3.2$ Mpc, the X-ray luminosity of HoII X-1 corresponds to the Eddington mass of $\approx$ 80 M$_{\sun}$,
which is considered a rather strict lower limit to the mass of accreted compact
object if the X-ray emission is isotropic.

A more accurate distance to HoII of 3.39 Mpc was recently determined by Karachentsev et al. (\cite{Kar02}). The new distance estimate would result in only a minimal  difference in the size, flux, and luminosity values we have calculated assuming  a value of $D=3.2$ Mpc.

The ROSAT HRI and PSPC data of HoII X-1 were presented in detail by Zezas et al. (\cite{Ze99}), revealing a point-like,
variable source (on scales of days and years) at the edge of the compact H II region \#70 (Hodge, Strobel \& Kennicut \cite{Ho94}).
The ROSAT PSPC spectrum was best described by either a steep power-law with 
$\Gamma=2.7$ or a thermal plasma with $kT\sim 0.8$ keV.
Miyaji, Lehmann \& Hasinger (\cite{Mi01}; hereafter MLH01) found that the 
ASCA spectrum extends to harder energies. The hard part of the spectrum  
is best fitted with a flatter power-law with $\Gamma\sim1.9$ and intrinsic 
absorption above that of our galaxy. A multi-colour disk black body model 
(MCD, Mitsuda et al. \cite{Mit84}) did not fit the ASCA spectrum of HoII 
X-1, unlike some other ULX's of similar luminosities. A joint 
PSPC-ASCA spectral analysis showed a soft excess component above 
the power law component. The soft excess could be described by either 
a MCD model with $kT_{in}=0.17$ keV, or a thin thermal plasma with 
$kT=0.3$ keV.  
MLH01 disfavored the MCD interpretation for the 
soft excess based on the large discrepancy between the black hole mass 
estimated from $kT_{in}$ and that estimated from the normalization. 
Furthermore, the spatial analysis of the ROSAT HRI image indicates an extended 
component.

In this paper we report on optical integral field and long-slit spectroscopic observations of the ultraluminous X-ray source X-1 in Holmberg II, as well as on results from the spatial and spectral analysis of the public {\it XMM-Newton} and {\it Chandra} data.
 
The outline of this paper is as follows. The optical imaging, and the long-slit and integral field spectroscopy of HoII X-1 are presented in 
Sect. 2. The H II region associated with the X-ray source is presented in Sect. 3. The optical properties (e.g., emission line flux maps, velocity dispersion, and radial velocities) of the H II region are described in Sect. 4. 
In Sect. 5 we present the X-ray spectral and spatial analysis based on public {\it XMM-Newton} EPIC-PN and {\it Chandra} ACIS-S data.
The implications of our results with respect to the nature of the ULX HoII X-1 are discussed in Sect. 6.

   \begin{table*}[t]
      \begin{center}
      \caption[]{Instruments and observations}
         \label{instr}
\begin{tabular}{|l|llll|}
\hline \noalign{\smallskip}
                    & \multicolumn{2}{c}{{\bf integral field spectroscopy}} & {\bf long-slit spectroscopy} & {\bf imaging}\\ 
\hline \noalign{\smallskip}
telescope           & Calar Alto 3.5m     & SAO 6m       & SAO 6m & CFHT 3.6m\\
instrument          & PMAS                 & MPFS          & LSS  & OSIS \\
\hline \noalign{\smallskip}
date of observation & 28.10.2001           & 15.03.2002    & 14-15.01.2002 & 11.03.2000\\
exposure time       & 4$\times$900s mosaic,- offsets: 4$^{\prime\prime}$       & 2$\times$900s  & 1800s (N1-3) & 2$\times$1200s (H$\alpha$) \\
                    &                 & & 2700s (N4-5)    &2$\times$300s ($R$) \\
                    &                 &           &   &2$\times$500s ($B$) \\
seeing              & 1.0$^{\prime\prime}$                &  1.5--2.0$^{\prime\prime}$        & 1.5$^{\prime\prime}$/1.1$^{\prime\prime}$ & $\sim0.7^{\prime\prime}$\\
field of view       & 8$^{\prime\prime}$$\times$8$^{\prime\prime}$, 12$^{\prime\prime}$$\times$12$^{\prime\prime}$ (mosaic)            & 16$^{\prime\prime}$$\times$16$^{\prime\prime}$     & 2$^{\prime\prime}\times140^{\prime\prime}$ & 3$^{\prime}$$\times$3$^{\prime}$\\
pixel scale         & 0.5$^{\prime\prime}$/pix             & 1.0$^{\prime\prime}$/pix      & 0.41$^{\prime\prime}$/pix& 0.088$^{\prime\prime}$/pix\\
image size in pixels& 2048$\times$4096 (binned 2$\times$2)    & 1024$\times$1024 & 1024$\times$1024 & 2048$\times$2048\\
spectral coverage   & 4450-5140 \AA        & 4210-6820 \AA & 4330-6750 \AA  & -\\
dispersion          & 0.76 \AA/binned pix   & 2.7 \AA/pix    & 2.4 \AA/pix & -\\
spectral resolution & 1.5 \AA              & 7.0 \AA        & 7.2 \AA &-\\
\noalign{\smallskip}
\hline
\end{tabular}
   \end{center}
   \end{table*}

\section{Observations and data reduction}

We have carried out integral field observations with the Potsdam Multi-Aperture Spectrophotometer (PMAS; Roth et al. \cite{Ro00}) and with the Multi-Pupil Fiber Spectrograph (MPFS; Afanasiev et al. \cite{Afa95a}), and long-slit observation with the Long--Slit Spectrograph (LSS; Afanasiev et al. \cite{Afa95b}) to determine the nature of the optical counterpart of HoII X-1. These observations are complementary due to the different technical properties of each instrument, e.g.; FOV, pixel scale, and spectral resolution, and due to the observational conditions (see Table \ref{instr}). Because the accurate {\it Chandra} position (see Sect. 5) was not know at the time of our observations, it was especially necessary to cover a large field of view with the integral field technique. Fig. \ref{FOV} shows the overlays of the FOV for each instrument on the CHFT archival $H\alpha$ image (see Sect 2.3) of the HoII X-1 region.  A detailed description of the observation is given below.

\subsection{Integral field spectroscopy with PMAS and MPFS}

We have used the PMAS at the Calar Alto 3.5m Telescope to obtain integral field observation of the optical counterpart of HoII X-1. 
The observations were part of a Science Verification run from October
23-28, 2001.  We have observed on Oct. 28
a set of 4 mosaic pointings, each offset by 4$\times$4~arsec in
four different directions from the previous deep field in order to search for spectral
signatures which could be associated with HoII X-1.

Motivated by the detection of the He II emission at the edge of the PMAS mosaic FOV (see Sect. 4.1) 
we conducted further observations with MPFS
at the 6-m SAO telescope in Russia. The LSS observations, obtained before the MPFS data (see Table \ref{instr}), provided valuable insight as to point the MPFS instrument exactly on the He II region.
The ''heel'' of the foot--like H II region HSK \#70 (Hodge, Strobel \& Kennicutt \cite{Ho94}) was centered on the CCD frame during the MPFS observations (Fig. \ref{FOV}).

The PMAS and MPFS data were reduced using P3d, an IDL based data reduction package developed at 
the Astrophysikalisches Institut Potsdam (Becker \cite{Be02}). The bias was subtracted using 
a bias exposure taken at the beginning of each night. Continuum lamp and Mercury (PMAS) or Neon (MPFS) emission line
lamp exposures were taken before and after the science exposures. 
The continuum lamp exposures were used to trace the individual spectra. 

The line lamp exposures were used for wavelength calibration.
The small number of mercury lines in the PMAS calibration spectra 
does not allow for a very accurate wavelength calibration. 
Therefore we have not derived radial velocities from the PMAS data. 

The fiber throughput variations were calibrated using a sky flat taken at the 
end of the night. The absolute flux calibratio of the PMAS and MPFS spectra was 
obtained using standard star exposure of HR153 and G191B2B, respectively, 
 just before the science exposures.

 The PMAS images were combined to a single mosaic frame after correcting for
 atmospheric refraction (Filippenko \cite{Fi82}), in each individual exposure.

\subsection{Long-slit observations}

Observations were carried out with the LSS instrument installed at the prime focus of the 6-m SAO telescope.
The slit orientation for the LSS spectra (N1-5) was along the main axis of the foot--like H II region, P.A.~$\approx
120^{\circ}$ (see Fig.\ref{FOV}).  The relative offsets between the slits N2 and N3, and between the slits
N3 and N4, are 0.6$^{\prime\prime}$ and 1.3$^{\prime\prime}$, respectively. The slit N1 was offset 3.3$^{\prime\prime}$ North-East of N2, and the slit N5
was offset 2.5$^{\prime\prime}$ South-West of N4.

The accurate locations of the slits were determined using images
from the TV--guiding camera on the LSS spectrograph. The FOV of the
guiding camera is 2$^{\prime}$. Five stars surrounding the target were used
to determine the coordinates of the slit in each image.

We have use standard MIDAS procedures to reduce the LSS spectra.
The wavelength calibration was checked again using the [O I]\,$\lambda 5577$ and
$\lambda 6300$ sky lines.

   \begin{figure}[t]
    \hspace*{1.0cm}\psfig{file=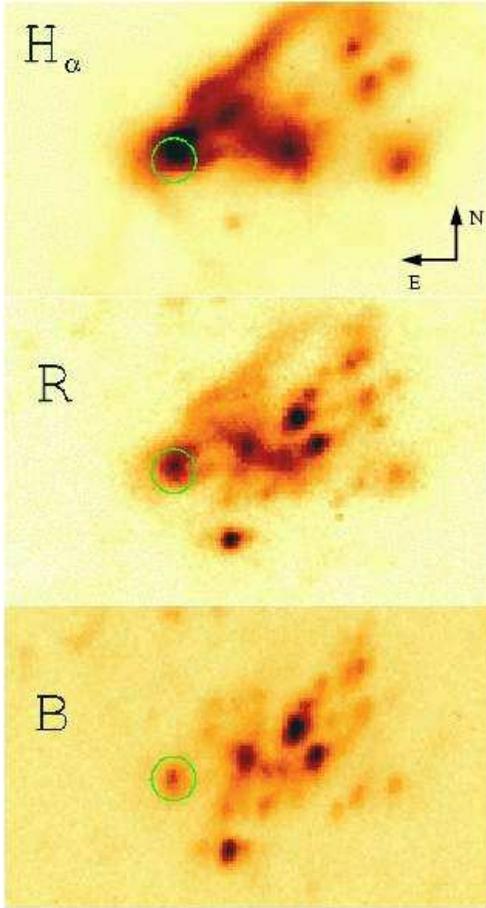,width=6.5cm,clip=}
      \caption{CFHT $H\alpha, R$, and $B$ images of the HoII X--1 region with the marked {\it Chandra} position (a circle 1" in radius) of the X-ray source. The image size is 24.5" $\times$ 14.0"}
         \label{CFHT}
   \end{figure}

   \begin{figure}[t]
   \centering
   \psfig{file=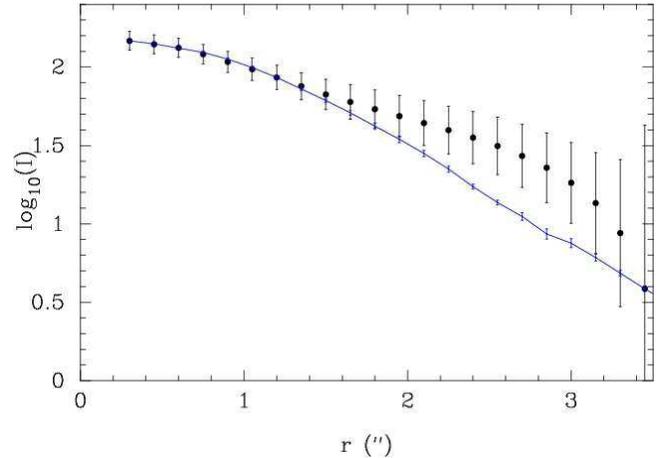,width=8.5cm,clip=}
      \caption{
Surface brightness profile from the He II emission around the ULX (solid
circles) together with the MPFS PSF (solid line). The error bars includes
both the photon noise and the error in the surface brightness
determination. The He II region around the ULX is clearly extended to
about 3$^{\prime\prime}$.
}
         \label{profile}
   \end{figure}

\subsection{CFHT $BRH\alpha$ archival imaging data}

We adopted the accurate {\it Chandra} coordinates and applied 
linear astrometry to the CFHT  $H\alpha$, $R$, and $B$ archive images
to determine the position of the X-ray source in the H II region.

For an absolute photometric calibration two images of the globular cluster
NGC4147 taken during the same night were utilised. 
Six standard stars from the cluster (Odewahn et al.  \cite{Ode92}) were used to calibrate the images. 
The calibration was double checked by independent measurements of the comparison stars 
in B-band CCD images made by V. Goranskii with the 1-m SAO telescope.

\section{Optical counterpart of the ULX HoII X-1}

To obtain a reliable astrometry of the CFHT $H\alpha, R$, and $B$ archival images we 
used the USNO-B1.0 catalogue. The astrometric accuracy of the catalogue is about 0.2$^{\prime\prime}$ at J2000. 
About fifteen USNO-B1.0 stars were found within the OSIS FOV around HoII X-1.
The standard deviations of the star positions are 
0.20$^{\prime\prime}$ for the $H\alpha$, 0.13$^{\prime\prime}$ for the $B$, and 0.15" for the  $R$ images.
In Fig. \ref{CFHT} we present the CFHT images with
the {\it Chandra} ACIS-S source position marked (see detailed description in Sect. 5).
An object on the $B$ image perfectly coincides with the position of the X--ray source.
Its magnitude is $B=20.5\pm0.1$ mag.

Since all the stars on the $B$ image in Fig. \ref{CFHT} look elongated, while
the seeing was fairly good ($\approx 0.7$"), we fitted 2d-Gaussians to
23 stars surrounding the X-ray position.
The elongation of these stars was $b/a = 1.36 \pm 0.05$, where the major axis size is $b = 0.89 \pm 0.04$" and  the minor axis is $a = 0.65 \pm 0.01$". The
position angle of the star images was P.A.$=-19.2 \pm 3.0^{\circ}$.

The object coincident with the X-ray source has the following parameters: 
$b/a \approx 1.29$, $b \approx 1.22$", $a \approx 0.96$" and P.A.$ \approx -10^{\circ}$, which indicates that in comparison to the surrounding stars the counterpart is an extended object, e.g. a nebula or a compact stellar cluster. Its intrinsic size is about 0.69$^{\prime\prime}$ (11~pc) in the West-East direction and about 0.85$^{\prime\prime}$ (13~pc) in the North-South direction. The orientation is P.A.$\sim 0^{\circ}$.

Assuming the $B$-band flux calibrations from Allen (\cite{All73}), and a line-of-sight extinction to HoII of $A_B = 0.14^m$ (Schlegel et al. \cite{Schl98}), the optical luminosity of the counterpart is estimated to be $L_B \sim 6 \cdot
10^{37}$~erg/s. The absolute magnitude of this object is $M_B = -7.2$. 
Using the X-ray luminosity in the 0.3-8.0 keV energy band of $L_x \sim 10^{40}$~erg/s, corrected for absorption (see Sect. 5.2), we find $L_x/L_B \ge 170$. Recent {\it HST} ACS observations of HoII X-1, (published after the submission of this paper), resolved this object into several young stars (Kaaret et al. \cite{Kaa04}), which agrees with our interpretation. Due to the superb angular resolution of the ACS images Kaaret found a bright, point-like optical counterpart consistent either with a star with spectral type between O4V and B3 Ib, or reprocessed emission from an X-ray illuminated accretion disk. Because the star found by Kaaret et al. (\cite{Kaa04}) is about one magnitude fainter compared to our extended blue counterpart this results in a $L_x/L_B$ of $\sim300-400$. 

The MPFS He II $\lambda4686$ line flux map (see Fig. \ref{mpfs_flux1} in Sect. 4.1) clearly shows a He II emission line region at the {\it Chandra} ACIS-S position, and coincident with the optical counterpart detected on the CHFT $B$ image. This confirms the classification of ULX HoII X-1 as an X-ray ionized nebula (Pakull \& Mirioni \cite{Pak01}).
In order to determine if the He II emission is extended we have
derived the surface brightness profile of the He II region, using an algorithm
based on Jedrzejewski (\cite{Jed87}).  This method increases the signal-to-noise in the outer part of the surface brightness profile, since it comprises an average of the brightness along the eccentric anomaly. A similar technique is
extensively used for the detection of host galaxies in QSOs
(eg., S\'anchez \& Gonzalez-Serrano \cite{San03}).

The MPFS PSF was built using the continuum emission at a wavelength range
near the He II emission. Fig. \ref{profile}
shows the surface brightness profile of the He II region together with the surface brightness profile of the PSF, scaled to the peak of the He II emission. The He II emission is clearly extended beyond r$\sim$2$^{\prime\prime}$, which is also confirmed by the {\it HST} ACS He II narrow band image of Kaaret et al. (\cite{Kaa04}).

We find an extended and elongated He II $\lambda4686$ region with nearly the  same positional angle of P.A.$\sim-19^{\circ}$ as found for the blue counterpart on the $B$ image. The size of the He II $\lambda4686$ region after PSF correction is about 1.4 $\times$ 3.0$^{\prime\prime}$ (see Fig. \ref{profile}), which corresponds to about 21 $\times$ 47 pc. The larger size of the He II region compared with the blue counterpart (11$\times$14 pc) could be considered as an argument for a stellar complex, where the $B$-band counterpart represents the continuum emission and the He II region represents the high excitation nebula. 

The most important observational results from this section are that the He II emission is: extended, and in the same positional angle as the extended blue counterpart, and centered on the X-ray source. This confirms the classification of HoII X-1 as an X-ray ionized nebula as suggested by (Pakull \& Mirioni \cite{Pak01}).

   \begin{figure}[t]
   \begin{center}
\hspace*{0.1cm}   \psfig{file=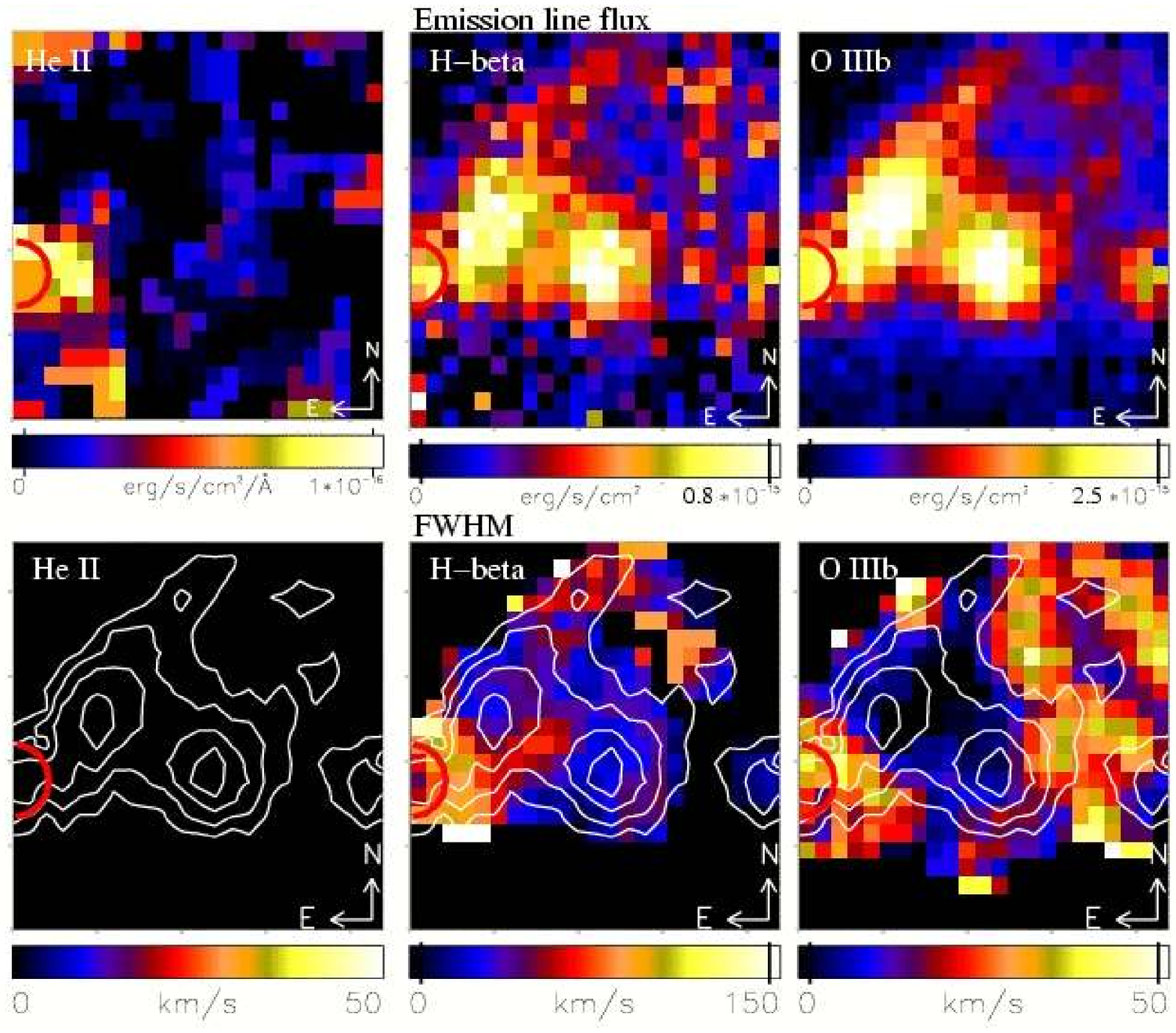,bbllx=205pt,bblly=23pt,bburx=574pt,bbury=494pt,width=8.5cm,clip=}
   \end{center}
      \caption{PMAS emission line flux maps (upper panels), and emission line FWHM maps (lower panels), corrected for instrumental resolution, for H$\beta$ and for [O III]~$\lambda5007$ in the HolI X-1 region. The FWHM images have been smoothed using a 3 $\times$ 3 pixel box. The emission line flux contour of [O III] $\lambda5007$ is overlayed on the FHWM maps. The ACIS-S error circle of the X-ray source position is shown. The FOV of the single images is about 11 $\times$ 11.5$^{\prime\prime}$. The black color in the flux maps and in the FWHM maps mark regions with no significant line detections and unresolved lines, respectively. The white color in the same maps indicate values above the limit given in the color bars.}
         \label{pmas_flux}
   \end{figure}

   \begin{figure}[t]
   \centering
   \psfig{file=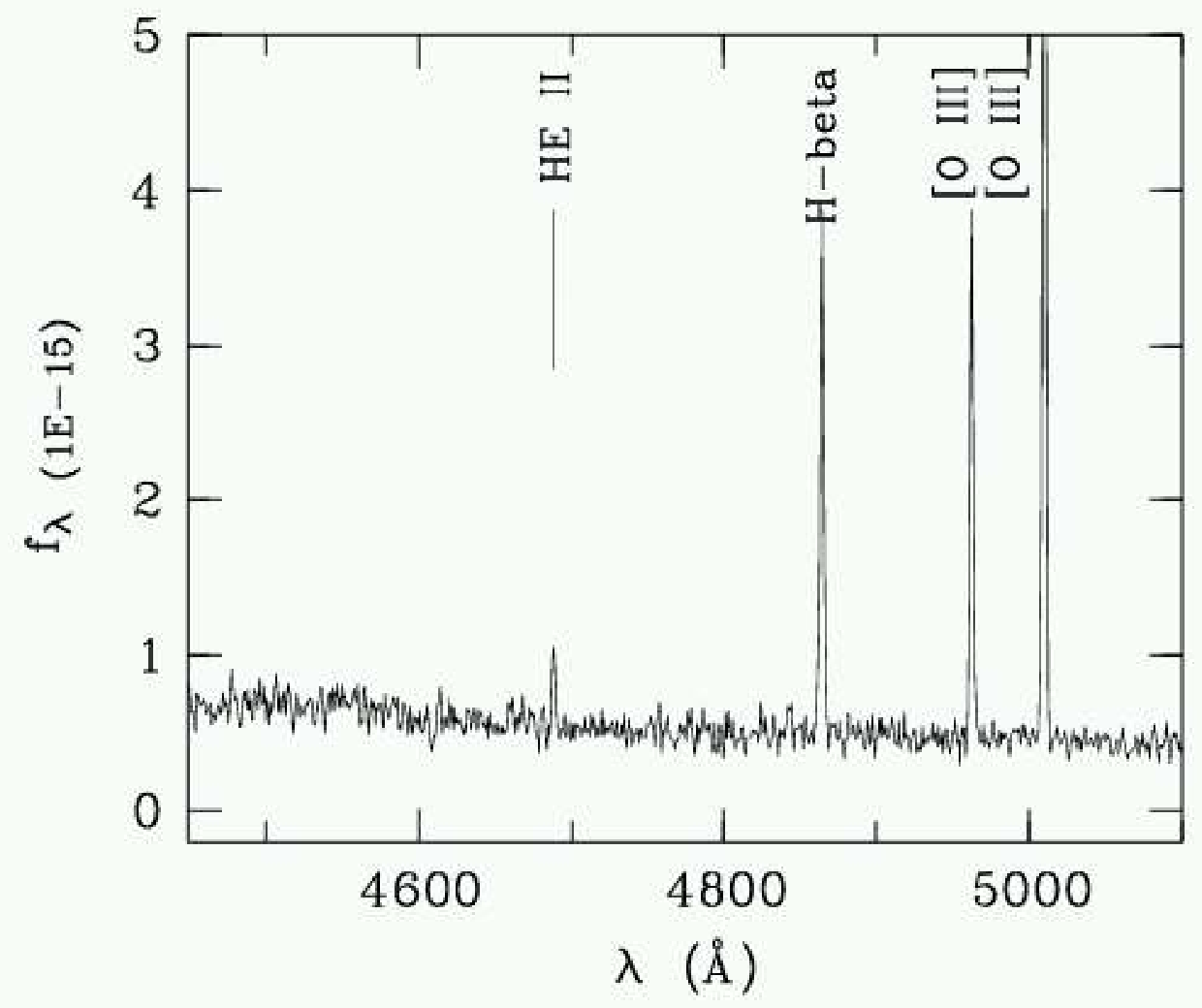,bbllx=50pt,bblly=95pt,bburx=405pt,bbury=396pt,width=8.5cm,clip=}
      \caption{The co-added PMAS spectrum at the X-ray position (see Fig. \ref{pmas_flux}) clearly shows an He II $\lambda4686$ emission line.}
         \label{pmas_spec}
   \end{figure}

\section{Emission line properties}

To determine the physical parameters (e.g. the radial velocities and the velocity dispersions) of
the H II region associated with the ULX, we have measured the emission line properties of all lines in the LSS, PMAS and MPFS spectra. Whereas the PMAS spectra cover only the He II $\lambda4686$, H$\beta~\lambda 4861$ and the [O III] $\lambda\lambda 4959, 5007$ emission lines, the MPFS spectral range includes
the H$\gamma~\lambda4363$ and H$\alpha~\lambda6563$ emission line regions as well.

Each emission line in the 240 MPFS and 506 PMAS integral field spectra was fitted with a Gaussian profile applying the Levenberg-Marquardt algorithm (Press et al. \cite{Pre92}). The four adjustable parameters were the total line flux, the mean wavelength, the sigma width of the Gaussian, and the flux of the local linear continuum. The Levenberg-Marquardt algorithm also estimates one-sigma errors
for each parameter. We considered only the emission line parameters of those lines which were at least 5$\sigma$ detections in the PMAS spectra, or 3$\sigma$ in the MPFS spectra. We have visually checked the reliability of the chosen line detection thresholds for the each spectrum.

\subsection{Line fluxes and He II line luminosity}

   \begin{figure*}
   \centering
   \psfig{file=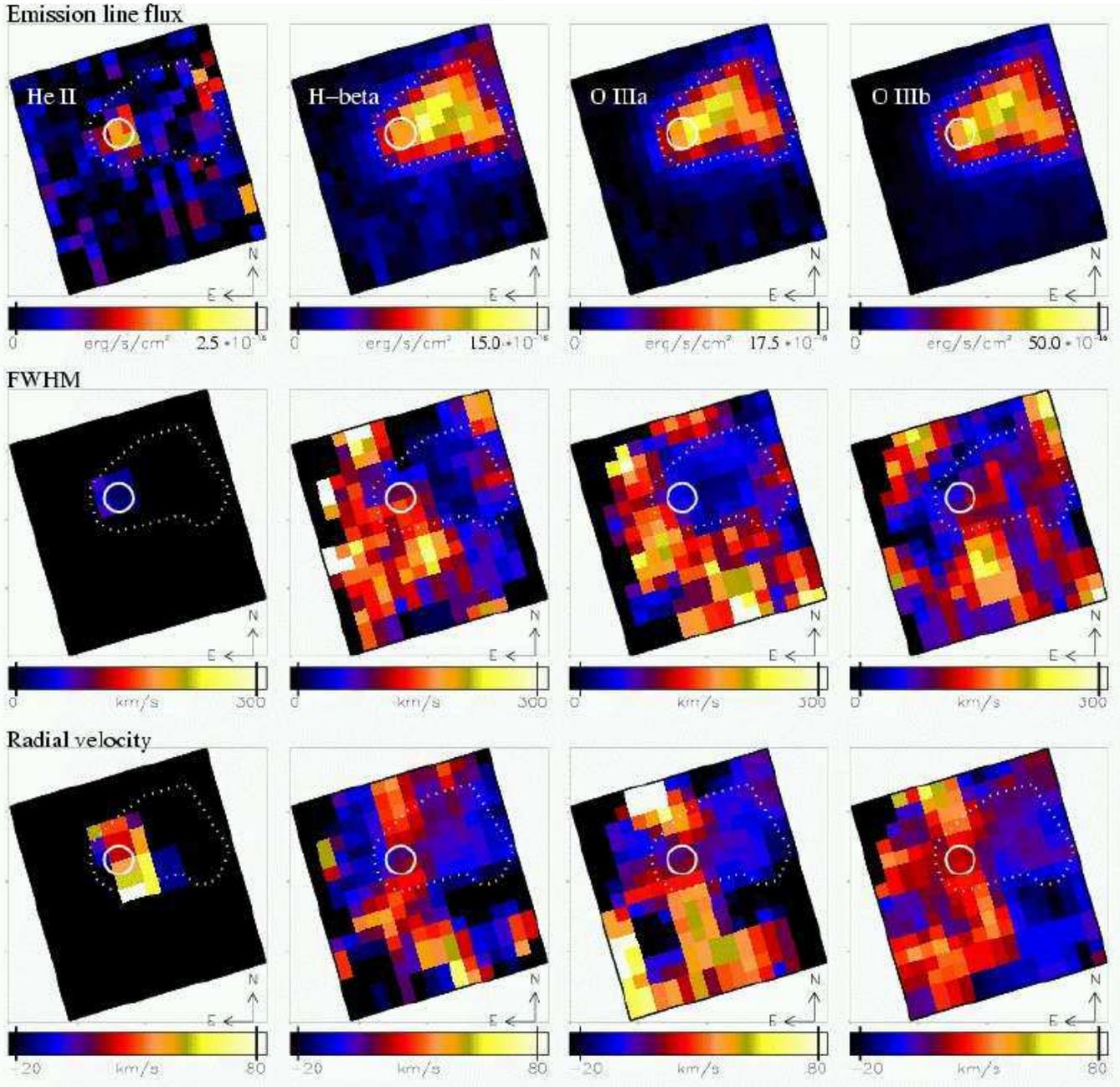,width=18.0cm,clip=}
      \caption{MPFS emission line flux, FWHM (corrected for instrumental resolution) and radial velocity maps for He II, H$\beta$, [O~III] $\lambda4959$ and [O~III] $\lambda5007$ of the HoII X-1 region. The black color in the flux/radial velocity maps and in the FWHM maps marks regions with no significant line detection and unresolved lines, respectively. The ACIS-S error circle of the X-ray source position is marked. The FOV of the single images is about 15$^{\prime\prime}\times 16^{\prime\prime}$. The dotted line in each map marks the area of the H II region given by the [OIII] $\lambda5007$ line flux larger than $\sim17 \times10^{-16}$ erg s$^{-1}$ cm$^{-2}$. The FWHM and radial velocity images have been smoothed using the mean of a 3 $\times$ 3 pixel box.}
         \label{mpfs_flux1}
   \end{figure*}

\begin{figure}[t]
   \centering
   \psfig{file=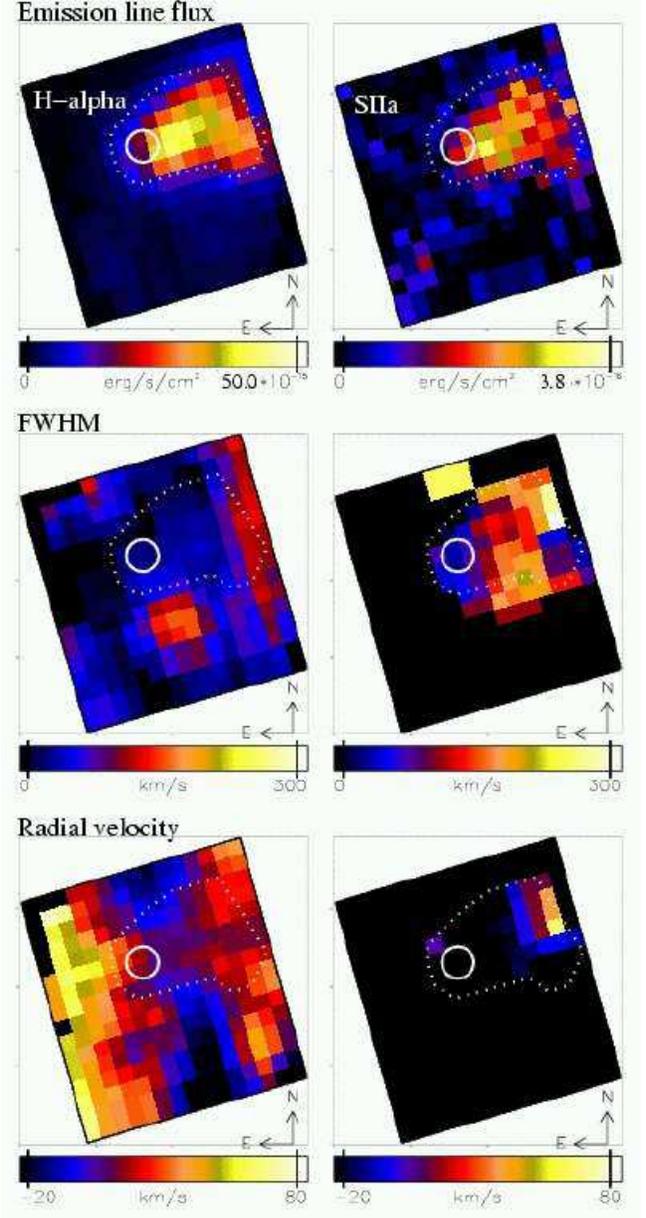,width=8.5cm,clip=}
      \caption{Same maps as shown in Fig. \ref{mpfs_flux1} for H$\alpha$ and [S II] $\lambda6717$.}
         \label{mpfs_flux2}
   \end{figure}

The emission line flux maps of H$\beta$ and [O III] $\lambda\lambda 4959, 5007$ derived from the PMAS mosaic spectra, and of He II, H$\beta$, [O III] $\lambda\lambda 4959, 5007$, H$\alpha$ and [S II] $\lambda6716$ derived from the MPFS spectra, are presented in the upper panel of Fig. \ref{pmas_flux}, \ref{mpfs_flux1} and \ref{mpfs_flux2}. The linear flux scale given below the maps starts with the blue color. The black regions mark the non-detection of a 5 sigma line for PMAS spectra and of a 3 sigma line for MPFS spectra. The circle with 1$^{\prime\prime}$ radius marks the X-ray position determined from the Chandra ACIS-S data (see Sect. 5). 

Because the PMAS data were taken under better seeing conditions and due to the better angular sampling of PMAS (see Table \ref{instr}), the PMAS emission line flux maps of H$\beta$ and [O III] $\lambda5007$ show more details compared to the MPFS maps. For instance the peak of both lines in the MPFS data is clearly resolved into two clumps with PMAS. 

He II emission lines are clearly detected from a relatively compact region inside the H II region \#70 at the {\it Chandra ACIS-S} position (see Fig. \ref{mpfs_flux1}). 
He II emission lines are not detected in the individual PMAS spectra due to the short exposure time. However, the He II line is clearly seen in the co-added PMAS spectrum (see Fig. \ref{pmas_spec}). In addition, the co-added spectrum suggests of an increasing blue continuum at wavelength below 4700 \AA, which is probably produced by the young stars resolved with {\it HST}. The increasing continuum to the blue is seen as well in the LSS spectrum (see Fig. \ref{LSS}). The absolute continuum flux of the PMAS and LSS spectra are not comparable because of the different amount of sky background emission. Nevertheless, the increasing blue continuum cannot be explained by the sky background.

The fluxes of all but the He II $\lambda4686$ emission lines peak 
outside the X-ray error circle (see Fig. \ref{mpfs_flux1} and \ref{mpfs_flux2}).
The peak PMAS fluxes agree well with the peak MPFS fluxes. 

In Fig. \ref{LSS}  we present an averaged LSS  spectrum of N2 and N3 at the location of the blue extended
counterpart. This spectrum covers the brightest region in He\,II;  
however, the spectrum did not cover the region completely. 
The real flux could be a factor of $\sim1.5$ times larger.
The brightest lines in the spectrum are the hydrogen lines and
[O III]\,$\lambda\lambda 4959, 5007$. No notable absorption lines or broad wings in
the permitted lines are observed; however the faint blue continuum is clearly
seen. All significantly detected lines are narrow, and formed in the nebula.
He II $\lambda 4686$ emission is the strongest permitted line (after the Hydrogen
lines) indicating a high excitation of the X--ray ionizated nebula (XIN, Pakull \& Mirioni \cite{Pak01}).

Relative intensities of emission lines in the LSS spectra agree very well
with those from the MPFS/PMAS spectra, however they are slightly different
in individual LSS-spectra. The latter implies that the physical conditions
of the gas may be different in different parts of the XIN. The relative fluxes of the MPFS spectra and the averaged LSS spectra N2 and N3 in units of the H$\beta$ flux are given in Table \ref{relflux}.

From the He II\,$\lambda 4686$ flux map in Fig. \ref{mpfs_flux1} we have determined a total line flux of 1.55 $\times$ 10$^{-15}$ erg s$^{-1}$ cm$^{-2}$ corresponding to a luminosity of $\sim1.9\times 10^{36}$ erg s$^{-1}$.
Using the averaged LSS spectra N2 and N3 at the X-ray source position we have determined a total He II~$\lambda 4686$ luminosity of $L(He II) \approx 1.4 \cdot 10^{36}$~erg s$^{-1}$, where the He II equivalent width is 29.7~\AA.
Furthermore, we have estimated the total He II~$\lambda 4686$ flux from the monochromatic MPFS images in Fig. \ref{Ions}. Assuming an 5" $\times$ 5" area of the He II region (not PSF corrected) we can directly measure the mean flux of log f$_{He II}\sim-16.1$, which gives a luminosity of $L \approx 2.2 \cdot 10^{36}$~erg s$^{-1}$.  
The total H$\beta$ luminosity in our LSS data is L(H$\beta) \approx 9.5 \cdot 10^{36}$~erg/s.

The He II luminosity agrees with the results published by Pakull \& Mirioni (\cite{Pak01}) of L(He II)$ \approx 2.5 \cdot 10^{36}$~erg/s, and by Kaaret et al. (\cite{Kaa04}) of L(He II)$ \approx 2.7 \cdot 10^{36}$~erg/s  (keeping in mind that our long-slits do not cover the entire He II region).  

\subsection{Diagnostic emission line flux ratios}

\begin{figure*}
   \centering
  \psfig{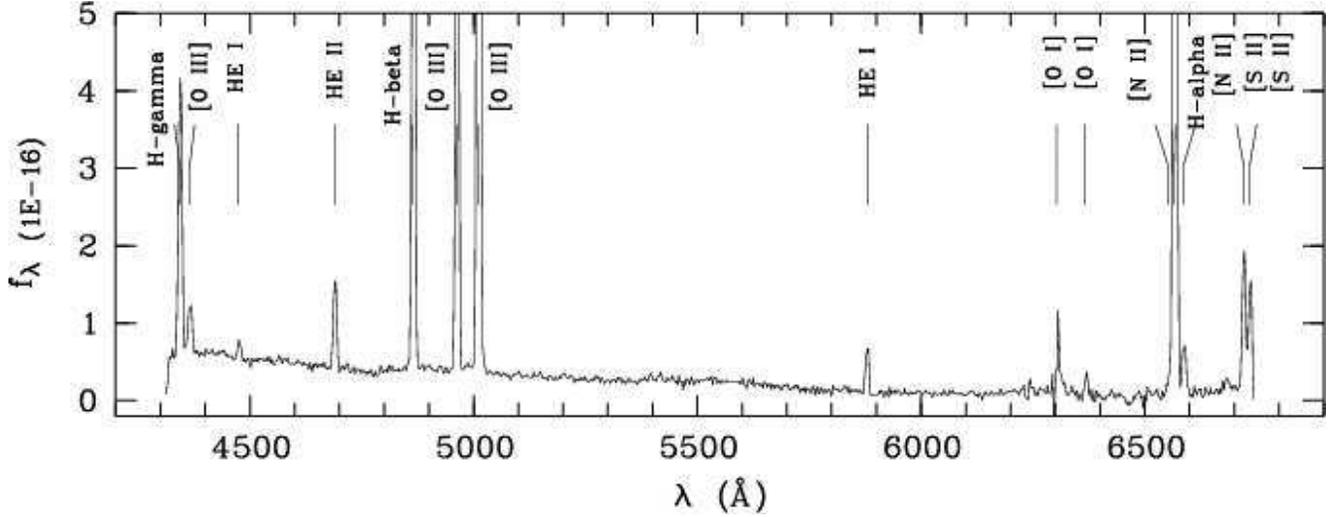}
      \caption{The average LSS spectrum of the slit positions 2 and 3 (see Fig. \ref{FOV}) at the X-ray source shows, in addition to the lines found
in the MPFS spectra, faint [O III] $\lambda4363$, He I $\lambda4471$ and [O I] $\lambda6364$ lines. The continuum emission seems to increase blueward $\sim$5700 \AA. }
         \label{LSS}
   \end{figure*}

\begin{figure*}
\hspace*{2cm}  \psfig{file=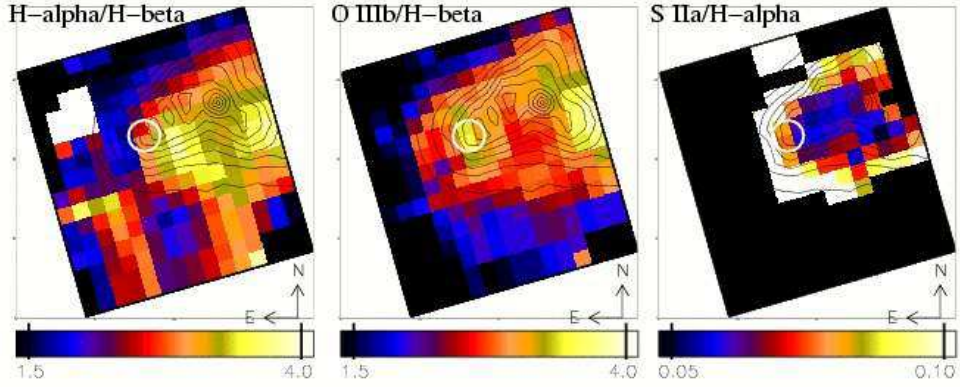,width=13.0cm,clip=}
     \caption{MPFS line flux ratio maps of Ho II X-1: H$\alpha$/H$\beta$, [O III] $\lambda5007$/H$\beta$ and [S II] $\lambda6717/$H$\alpha$. The [O III] $\lambda5007$ line flux contours are overlayed. }
        \label{flux_ratio}
  \end{figure*}

In order to understand the nature of the ionization inside the H II region \# 70 we determined the emission line flux ratios from the LSS and MPFS data.
Table \ref{ratioflux} gives the diagnostic ratios derived from the averaged LSS
at positions N2 and N3 (see Fig. \ref{LSS}). The MPFS line ratio maps of [O III] $\lambda5007$/H$\beta$, [S II] $\lambda6717$/H$\alpha$, and H$\alpha$/H$\beta$ are shown in Fig. \ref{flux_ratio}. 

The diagnostic line flux ratios ($[$O III$]~\lambda 5007$/H$\beta~ \lambda 4861$, $[$O I$]~\lambda 6300$/H$\alpha~ \lambda 6563$, $[$N II$]~\lambda 6583$/H$\alpha~ \lambda 6563$, and $[$S II$]~\lambda 6716+~\lambda 6731$/H$\alpha~ \lambda 6563$ see Table \ref{ratioflux}) derived from the averaged LSS spectrum and the MPFS line ratio maps agree well with the H II region classifications (Veilleux \& Osterbrock \cite{Vei87}). The $[$N II$]~\lambda 6583$/H$\alpha~ \lambda 6563$ ratio is at the lower end for H II regions (see Fig. 12.1 in  Osterbrock \cite{Ost89}). 
 
The line flux ratio of [S II] $\lambda6716+\lambda6731$/H$\alpha$ is well below 0.4 (see Table \ref{ratioflux}), which is an indication that emission comes from ionized gas in H II regions or in a nebula rather than from supernova remnants (Smith et al. \cite{Smi93}). 
 The 6, 20, and 90 cm radio data from Tongue \& Westpfahl \cite{Ton95}) shows a peak at 
the position: 8$^{h}$ 19$^{m}$ 28$^{2}$, $+$70$^{\deg}$ 42$^{\prime}$ 19$^{\prime\prime}$ (J2000), which is about 3$^{\prime\prime}$ West of the {\it Chandra} position (see Fig. \ref{FOV}). However, due to the low angular resolution of the data ($\sim15^{\prime\prime}$) the radio position is not precise enought to exclude a coincidence with the X-ray source. While the steep radio spectral index of $\alpha_{LC}=-1.0\pm0.2$ (between the 6 and 20 cm wavelength bands) favours supernova remnants (SNR) as the source of the radio emission, the optical emission line ratios suggest that the region of the radio emission is instead a H II region.

The MPFS line ratio maps of [O III] $\lambda5007$/H$\beta$ and [S II] $\lambda6717$/H$\alpha$ in Fig. \ref{flux_ratio} show no large variation over the H II region. Inside the {\it Chandra} ACIS-S error circle  and inside a region $\sim$10$^{\prime\prime}$  North-West of that position the MPFS flux ratio map shows slightly larger values of the [O III] $\lambda5007$/H$\beta$ ratio ($\sim3.3$) compared to for the main parts of the H II region ($\sim2.9$), which could indicate a slightly larger ionizating level in these regions. 

The H$\alpha$/H$\beta$ flux ratio map corrected for differential refraction (Fig. \ref{flux_ratio}) shows a peak ($\sim3.6$) about 3$^{\prime\prime}$ South-West of the X-ray position and a second peak ($\sim3.6$)  about 10$^{\prime\prime}$ West of the {\it Chandra}  position. The remaining parts of the H II region have a ratio of $\sim$2.4--2.9. The second peak is positionally consistent with the peak in the [O III] $\lambda5007$/H$\beta$ flux map. Interestingly, the H II region \# 70 seems to be much more extended to the South indicated by the red features in the lower panel of the flux ratio map.  

To determine the intrinsic flux ratio of the H II regions we used the case B Balmer recombination decrement f$_{H\alpha}$/f$_{H\beta}=2.85$ for
$T=10^{4}$ K and N$_{e}=10^{4}$ cm$^{-3}$ (Brocklehurst \cite{Bro71}). Assuming that the observed H$\alpha$/H$\beta$ ratio inside the H II region
is due to extinction we obtain $E(B-V)=0.02$ for when the ratio is 2.9 and $E(B-V)=0.21$ when the ratio is 3.6. In this case the H II region would show
a larger extinction at about 3$^{\prime\prime}$ South-West of the X-ray position and at about $\sim$10$^{\prime\prime}$ North-West of the X-ray position.
However, the larger  H$\alpha$/H$\beta$ ratio in these regions could even be due to larger ionization.

  \begin{table}
  \centering
     \caption[]{Relative line fluxes in units of the H$\beta \lambda4861$ flux  derived from MPFS and LSS data. The fluxes of red forbidden lines in units of the H$\alpha$ flux are given in brackets. }
        \label{relflux}
        \begin{tabular}{lcc}
           \hline
           \noalign{\smallskip}
            line & \multicolumn{2}{c}{relative flux}  \\
                 & LSS & MPFS \\ 
           \noalign{\smallskip}
           \hline
           \noalign{\smallskip}
           H$\gamma$ $\lambda 4363$   & 0.45  &  -   \\ 
           $[$O III$]~  \lambda 4363$ & 0.08  &  -   \\ 
           He\,I\,$\lambda 4471$      & 0.08  &  -   \\
           He\,II\,$\lambda 4686$     & 0.14  & 0.12 \\
           He\,I\,$\lambda 4713$      & 0.01  &  -   \\
           He\,I\,$\lambda 4922$      & 0.01  &  -   \\
           $[$O III$]~ \lambda 4959$  & 1.00  & 1.20 \\
           $[$O III$]~ \lambda 5007$  & 3.00  & 3.00 \\ 
           He\,II\,$\lambda 5412$     & 0.02  &  -   \\
           He\,I\,$\lambda 5876$      & 0.07  &  -   \\
           $[$O I$]$~$\lambda 6300$   & 0.11 (0.03) &  -   \\
           H$\alpha$ $\lambda 6563$   & 3.20  & 3.80 \\
           $[$N II$]$~$\lambda 6583$  & 0.08 (0.02) &  -   \\
           He\,I\,$\lambda 6678$      & 0.03  &  -   \\
           $[$S II$]$~$\lambda 6717$  & 0.22 (0.07) & 0.28 \\
           \noalign{\smallskip}
           \hline
        \end{tabular}
  \end{table}

  \begin{table}
  \centering
     \caption[]{Diagnostic emission line ratios from the LSS spectra in logarithmic units.}
        \label{ratioflux}
        \begin{tabular}{llc}
           \hline
           \noalign{\smallskip}
              nr.  &  ratio & value \\
           \noalign{\smallskip}
           \hline
           \noalign{\smallskip}
           1 & $[$O III$]~\lambda 5007$/H$\beta~ \lambda 4861$   & 0.48   \\ 
           2 & $[$O I$]~\lambda 6300$/H$\alpha~ \lambda 6563$   & -1.52  \\ 
           3 & $[$N II$]~\lambda 6583$/H$\alpha~ \lambda 6563$   & -1.69  \\ 
           4 & $[$S II$]~\lambda 6716+~\lambda 6731$/H$\alpha~ \lambda 6563$   & -0.93  \\ 
           5 & $[$S II$]~\lambda 6716$/$[$S II$]~\lambda 6731$  &  0.10  \\ 
           6 & $[$O III$]~\lambda 4959+~\lambda 5007$/$[$O III$]~\lambda 4363$  &  $\sim2$  \\ 
           7 & H$\alpha~ \lambda 6563$/H$\beta~ \lambda 4861$   &  0.51  \\ 
           \noalign{\smallskip}
           \hline
        \end{tabular}
\begin{list}{}{}
\item[]
\item[]
\end{list}
  \end{table}

Following Osterbrock (\cite{Ost89}) the line flux ratio [S II] $\lambda6716$/[S II] $\lambda6731$ was used to determine the electron density of region \# 70, which is about 200 cm$^{-3}$. This value is typical for H II regions. 
The temperature of the H II region of T$\sim$15000 K has been estimated from the line flux ratio [O III] $\lambda4959+\lambda5007$/[O III] $\lambda4363$ (Osterbrock \cite{Ost89}). 

The main result of this section is that the emission line flux ratios inside the Holmberg II region \#70, even at the positions of the X-ray and radio sources, are consistent with H II regions. 

\subsection{FWHM of emission lines}

In order to search for dynamical signatures of a black hole we determined the FWHM of the emission lines. The main problem is to correct the observed FWHM for instrumental resolution, which depends on the single pixel and the wavelength positions. We have derived the instrumentally corrected FWHM using the FWHM of the night sky lines and the FWHM of the emission lines from calibration lamp exposures. However, we cannot account for all instrumental effects, and these becomes especially significant for the lower resolution MPFS data.

  \begin{figure}[t]
  \hspace*{2cm} \psfig{file=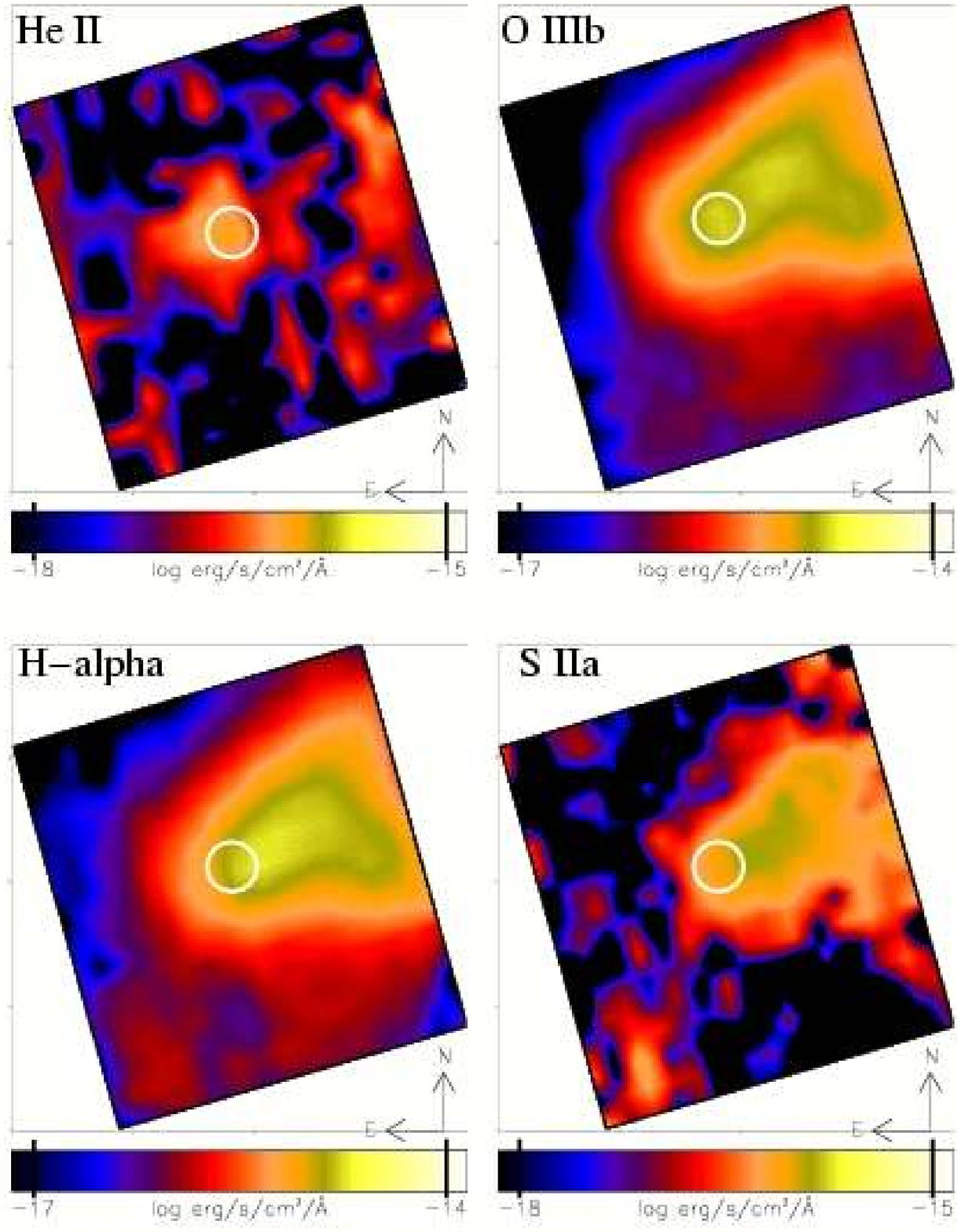,bbllx=31pt,bblly=448pt,bburx=302pt,bbury=786pt,width=4.5cm,clip=}
        \caption{Monochromatic MPFS images in He II $\lambda4686$
               smoothed with a 2 $\times$ 2 pixel kernel. The flux is given in a logarithmic scale. The Chandra ACIS-S source position
               is overlayed. The He II image shows a second, probably extended region in North-West direction of the X-ray source, similar to
               the He II emission line flux map (Fig. \ref{mpfs_flux1}), and the line intensity plots from the long-slit spectra.  }
        \label{Ions}
  \end{figure}

The PMAS FWHM maps and the MPFS FWHM maps (in km s$^{-1}$) are presented at the bottom of
Fig. \ref{pmas_flux} and in the middle of Fig. \ref{mpfs_flux1} and \ref{mpfs_flux2}, 
respectively. The color scale of the FWHM images was chosen to show reliable structures
of the velocity field. Due to the larger angular sampling (0.5$^{\prime\prime}$/pix) and the better spectral
resolution the PMAS spectra are better suited to derive the FWHM of the lines. Unfortunately,
the PMAS spectra show that the He II $\lambda4686$/$[$O III$]~\lambda5007$ emission region, and the PMAS mosaic do not cover the entire Holmberg II region \#70.

The lower panel of Fig. \ref{pmas_flux} shows that the PMAS FWHM of the H$\beta$ line inside the H II region is in general 
larger compared to that of the forbidden [O III]~$\lambda5007$ line. The PMAS FWHM of both lines peak around the position of the {\it Chandra} ACIS-S error circle. 

Because the PMAS and MPFS FWHM maps are given with different scales we present for comparison the peak values of the FWHM inside the H II region derived from both intruments in Table \ref{peakFWHM}.
The MPFS FWHM peak values are in agreement with those values derived from PMAS (keeping in mind the lower angular and spectral resolution of the MPFS data), except for [S II] $\lambda6717$, which is probably blended with the [S II] $\lambda6731$ emission line. The FWHM of H$\beta$ derived from PMAS reaches about 30 km s$^{-1}$ inside the H II region, but outside the He II sub-region, it is up to 80 km s$^{-1}$.  
The FWHM of the 30 km s$^{-1}$ inside the H II region (which translates into a velocity dispersion of about 13 km s$^{-1}$), is consistent with the velocity dispersion measurements of $\sigma_{H\alpha}=13.0\pm0.3$ km s$^{-1}$ and $\sigma_{[O III] \lambda5007}=11.3\pm0.2$ km s$^{-1}$ obtained by Hippelein (\cite{Hip86}). 

However,  the FHWM more than doubles at the X-ray position.
Assuming the  virial theorem and that the enclosed mass is related to the velocity change of $\sim$50 km s$^{-1}$ at the distance of $\sim$30 pc the resulting mass of a putative black hole would be about $10^{7}$M$_{\sun}$. Therefore we believe that the increased FWHM at the location of the He~II/X-ray source may be understood as dynamical influence of the putative black hole like the accretion disk wind or jets.

The MPFS FWHM maps show a very complex velocity field in the South-East direction of the {\it Chandra} position, but outside the main H II region. Furthermore there is a peak of $\sim150-200$ km s$^{-1}$ at about 5$^{\prime\prime}$ South-West of the X-ray source position in the MPFS FWHM maps of H$\beta~\lambda4861$, [O III] $\lambda5007$, and H$\alpha$. The complex velocity field outside the H II region is most probably not related to HoII X-1.

  \begin{table}
  \centering
     \caption[]{Peak values of the FWHM inside the compact H II region \#70 derived from the PMAS and MPFS data}
        \label{peakFWHM}
        \begin{tabular}{lcc}
           \hline
           \noalign{\smallskip}
              emission line  & \multicolumn{2}{c}{FWHM$_{peak}$}\\
           \noalign{\smallskip}
                 & PMAS & MPFS \\ 
           \hline
           \noalign{\smallskip}
           He II ~$\lambda4686$     &-                      & $\sim30$ km s$^{-1}$ \\ 
           H$\beta~ \lambda 4861$   & $\sim80$ (30$^{\ast}$) km s$^{-1}$ & $\sim80$ km s$^{-1}$ \\ 
           $[$O III$]~\lambda 4959$ &-                      & $\sim30$ km s$^{-1}$  \\ 
           $[$O III$]~\lambda 5007$ & $\sim30$ km s$^{-1}$ & $\sim80$ km s$^{-1}$\\ 
           H$\alpha~ \lambda 6563$  &-                      & $\sim30$ km s$^{-1}$ \\ 
           $[$S II$]~\lambda 6716$  &-                      & $\sim200$ km s$^{-1}$$^{\ast \ast}$\\ 
           \noalign{\smallskip}
           \hline
        \end{tabular}
\begin{list}{}{}
\item[$^{\ast}$ Peak value inside the H II but outside the He II region.]
\item[$^{\ast \ast}$ Probably blended with [S II] $\lambda6731$.]
\end{list}
  \end{table}

\subsection{Radial velocities}

   \begin{figure}
   \hspace*{1.7cm}   \psfig{file=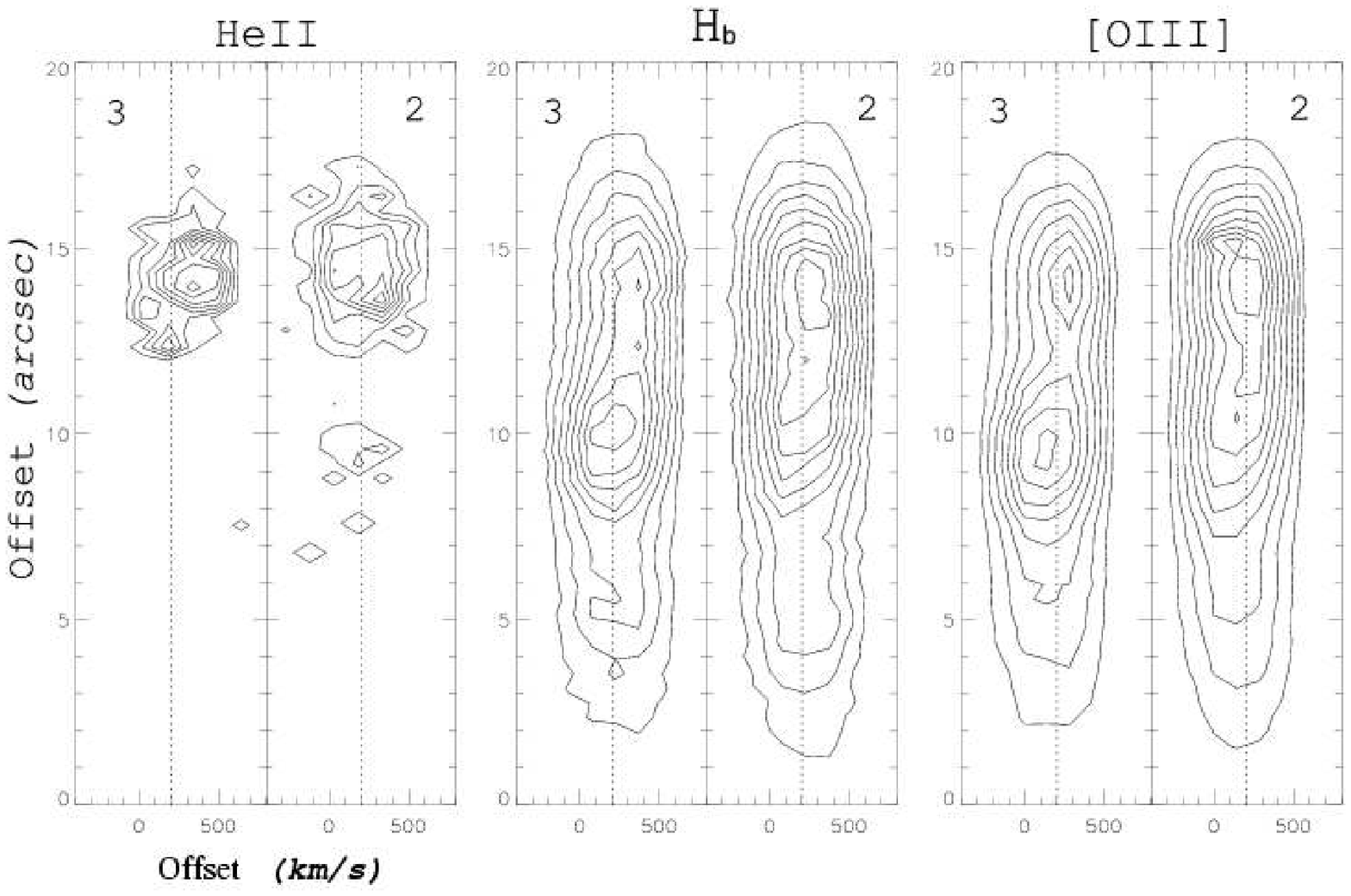,bbllx=21pt,bblly=21pt,bburx=217pt,bbury=386pt,height=9cm,clip=}
      \caption{
Isophotes of the 2d-LSS spectra at slit positions N2 and N3 in the He\,II\,$\lambda 4686$ emission line.  The extended He II region coincident with the X-ray source HoII X-1 is located at an offset value of $\sim14.5$$^{\prime\prime}$. A second fainter He II region is detected in the spectrum N2 at an offset value of $\sim 9$$^{\prime\prime}$ in the North-West direction along the slit.} 
         \label{isoph}
   \end{figure}

To determine the radial velocity field around HoII X-1 we have used
 the LSS spectra at the slit positions N2 and N3 and the MPFS data.

In Fig. \ref{isoph} we show the continuum and background subtracted line
 isophotes of the 2d-LSS spectra in the He\,II\,$\lambda 4686$ 
emission line. The position of the emission line (or equivalently, the position of the line emitting region) along the wavelength direction (accros the slit) is given as an offset in km s$^{-1}$. The position along the slit is given as an offset in arcsec, where East edge of the ''heel'' of the compact H II region \#70 is located at an offset of $\sim17^{\prime\prime}$ and the He II region at an offset of $\sim14.5^{\prime\prime}$. The offset values decrease in the North-West direction. 
Both spectra N2 and N3 cover HoII X-1. The He\,II peak intensity in N2  is 1.4 times greater than that in N3.

\begin{figure}[t]
   \centering
   \psfig{file=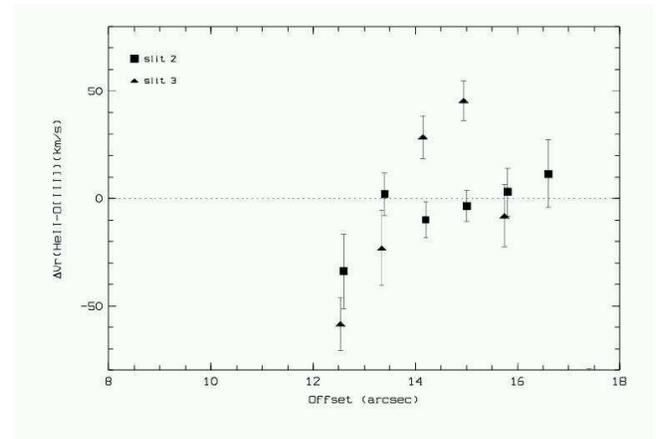,width=8.5cm,clip=}
      \caption{Radial velocities difference derived from the HeII~$\lambda4686$ and the [O III]~$\lambda5005$ emission lines inside the He II region along the slit position for the N2 and N3. The offset (arcsec) is the same as in Fig. \ref{isoph}, its bigger values correspond to the ''heel'' of the HII region.
}
         \label{vel_int}
   \end{figure}

The He\,II region is located near the edge of the H\,II region.
Along the slits we can directly study the  structure of the region and
we can see the complex structure of the He\,II emission. Even the 
shift in the slit position from N2 to N3 (0.6$^{\prime\prime}$) results in a
notable change of the isophote structures.

The line isophote structures across the slits can not be directly interpreted
as radial velocity variations, because the slit width (2$^{\prime\prime}$) is larger than the seeing.
In the isophotes shown in Fig. \ref{isoph}, the redshift corresponds to a shift 
of an emission knot covered by the slit to North-East direction.

There is an additional He\,II emission region, which is located about 6--7$^{\prime\prime}$ ($\sim 100$~pc)
in North-West direction of the  the He\,II/X--ray source (see Fig.  \ref{mpfs_flux1} and \ref{Ions}). The same He II emission region is detected in the spectrum N2.
The CFHT archival $B$ image in Fig. \ref{CFHT} shows that this is very complex region.
It may be that the He\,II emission is radiated by Wolf-Rayet stars.

In considering the radial velocity distributions in Fig. \ref{isoph} we have to remember
that the absolute values of the velocities are not correct as they depend on
the emission knot location within the slit. The complex structure of the radial velocity field
is more obvious in the MPFS radial velocity maps at the bottom panel of Fig. \ref{mpfs_flux1}
and \ref{mpfs_flux2}. The radial velocity maps of the strong emission lines, H$\beta$, 
[O III]~$\lambda4959$ and [O III]~$\lambda5007$ (except of H$\alpha$), show nearly the same 
structure in the H II region and its enviroment. The radial velocity ranges from about -20 
to 40 km s$^{-1}$ inside the H II region, and
reaches its maximum around the position of the X-ray source. 
Crossing the {\it Chandra} position nearly in East-West direction (see the maps of H$\beta$ to [O III]~$\lambda5007$ in Fig. \ref{mpfs_flux1})
the radial velocity changes from negative to positive to negative. The behaviour is
comfirmed by the radial velocities derived from the LSS spectra N2 and N3. 
If we assume a line crossing the {\it Chandra} position im North-South direction the radial velocity
shows only positive values. 

As already found for the MPFS spectra, the H$\beta$ and [OIII]~$\lambda5007$  radial velocities derived from the LSS spectra are absolutely the same. Because the LSS spectra have higher S/N we can study the radial velocity field much better. The LSS spectra show that there is a difference in the He\,II~$\lambda4686$ and [O III]~$\lambda5007$ radial velocities inside the He\,II region (see Fig. \ref{vel_int}).
Moving along the slit N2 from the
''heel's'' edge (offset position $\sim16-17^{\prime\prime}$)) toward the main nebula in North-West direction, the He\,II~$\lambda4686$ line shows about
the same velocity as the [O III]~$\lambda5007$ line, but it becomes negative (--30~km s$^{-1}$) at the offset position of 12--13$^{\prime\prime}$. Along the slit N3 the He\,II behavior is more complex. Its relative radial velocity is about zero at the
`heel's'' edge, positive (up to $+45$~km/s) in the middle of the
He\,II region, and negative (--50~km/s) at the offset position of
12--13.5$^{\prime\prime}$ (North-West of the ''heel''). The maximal differences
in the radial velocity of the He\,II~$\lambda4686$ line compared to the [O III]~$\lambda5007$ line is $\pm 50$~km s$^{-1}$ on spatial scales  of $\pm 2^{\prime\prime}$ ($\pm 30$~pc in projection) .

There is a complex radial velocity structure, which may be related to the X-ray source ionizing the surrounding gas. In this case the putative black hole not only ionize the surrounding gas, but also perturb the gas dynamically.
 
   \begin{figure*}
     \begin{center}
     \begin{minipage}{15.5cm}
      \begin{minipage}{9.5cm}
\hspace*{-1.0cm}   \psfig{file=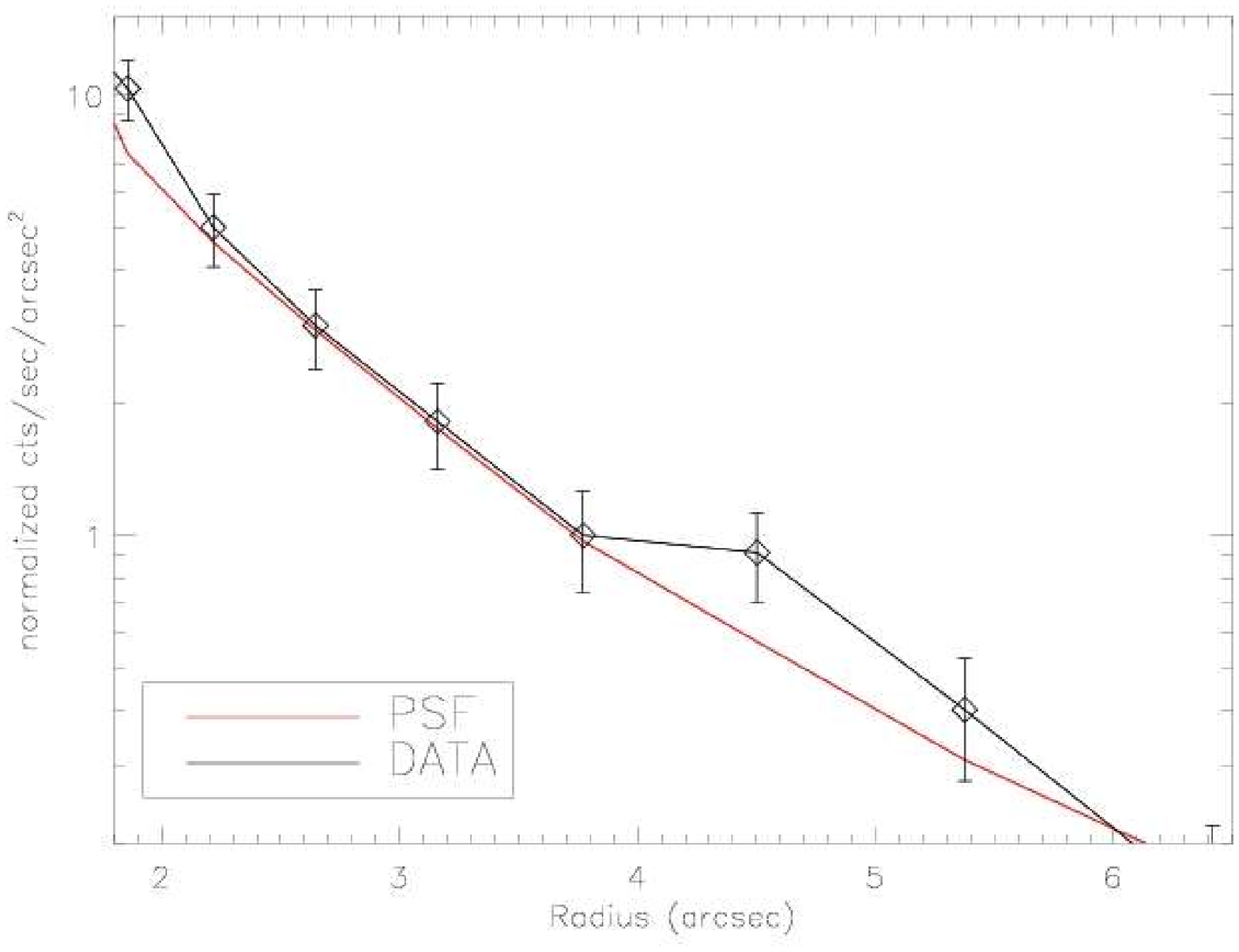,width=9.5cm,clip=}
      \end{minipage}
      \hfill
      \begin{minipage}{6.0cm}
   \psfig{file=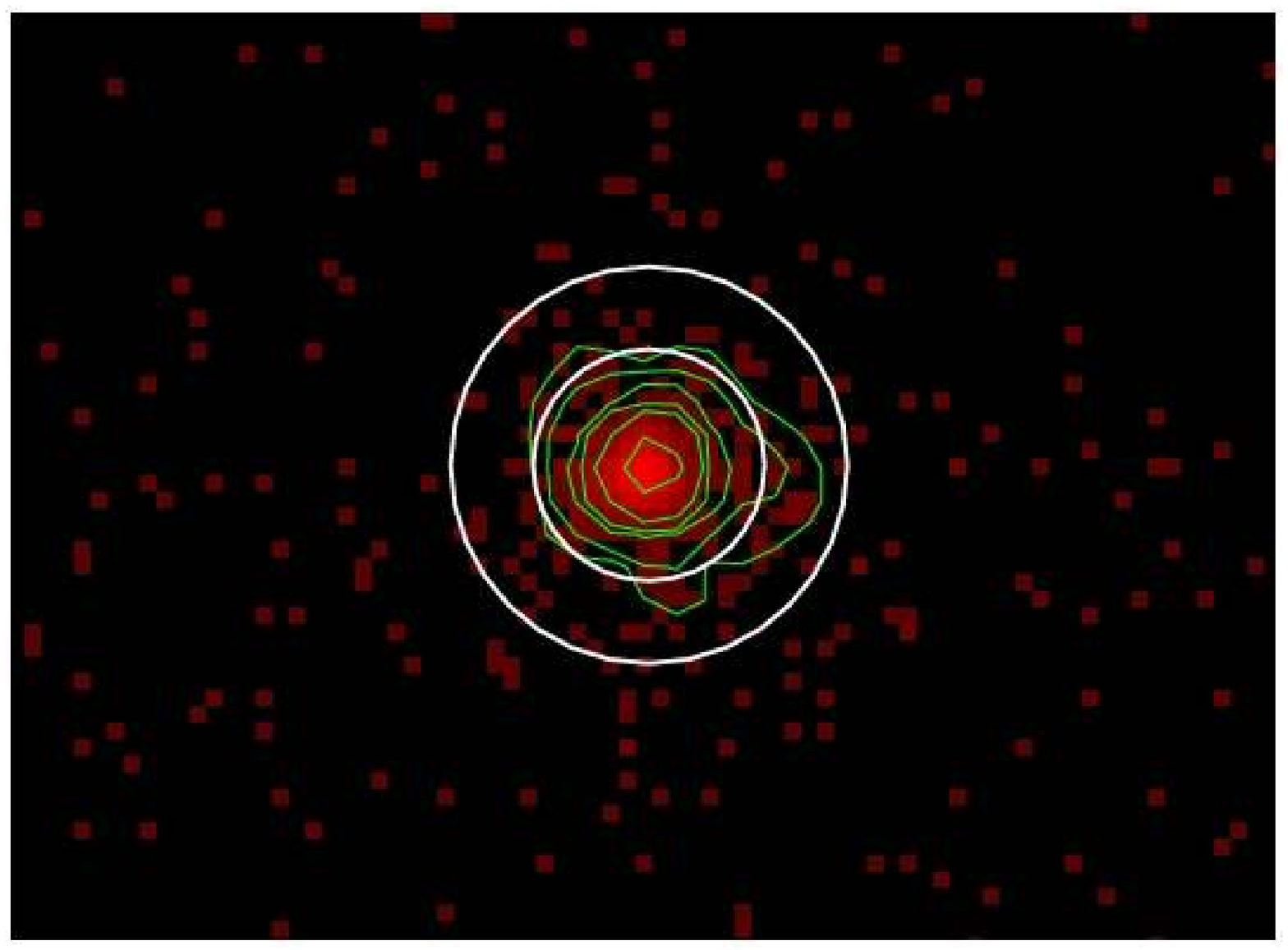,bbllx=172pt,bblly=104pt,bburx=423pt,bbury=354pt,width=6.0cm,clip=}
      \end{minipage}
     \end{minipage}
      \caption{
Left: Radial profile of Holmberg II X-1 (diamonds with error bars) compared with the simulated ACIS-S PSF at 1 keV. A pixel corresponds to 0.5$^{\prime\prime}$. The PSF was normalized to the maximum source counts at the first annulus outside 3$^{\prime\prime}$ (6 pixel) in radius. The counts in the radial range between $\sim$4-5$^{\prime\prime}$ (5. and 6. radial bins) are marginally above the 1 keV PSF. 
Right: Logarithmically scaled 5 ks {\it Chandra} ACIS-S image of HoII X-1 in the 0.3-8.0 keV energy band. The 5, 10, 30, 100, 300, 1000 and 3000 $\sigma$ contours above the logarithmic value of the background are shown (North is up, East is left). The FOV is about 17$^{\prime\prime}$$\times$17$^{\prime\prime}$. The circles with 7 and 12 pixel radius mark the annulus with slightly larger counts than expected from to the 1 keV ACIS-S PSF. 
}
         \label{xray_cont}
     \end{center}
   \end{figure*}

\section{Chandra and XMM observations of HoII X-1}

\begin{table}[b]
\footnotesize
\caption{Log of the analyzed {\it Chandra} and {\it XMM} Observations }
\begin{tabular}{ccc}
\hline\hline
Detector/Modes & Observation  & Exposure/Frame\\ 
\hline
{\it Chandra} ACIS-S  & Date: 02-NOV-2001 &  5.1 ks      \\
 Faint/Timed          & Seq. 600121       & 1/4 Subarray \\ 
{\it XMM} PN          & Date: 10-APR-2002 &  5.1 ks      \\
  Thin                & Seq. 0112520601   & PrimeFullWindow \\ 
{\it XMM} PN          & Date: 16-APR-2002 & 11.1 ks      \\
  Thin                & Seq. 0112522201   & PrimeFullWindow \\ 
{\it XMM} PN          & Date: 18-SEP-2002 &  5.0 ks      \\
  Thin                & Seq. 0112520901   & PrimeFullWindow \\ 
\hline
\end{tabular}
\label{tab:xlog}
\end{table}

\label{sec:intr}
 
 A 5.1 ks {\it Chandra} observation of this object was done
with the ACIS-S detector. The data were obtained from the public archive and a spectral and
spatial analysis was carried out. 
The log of the {\it Chandra} and {\it XMM} observations is shown in Table \ref{tab:xlog}.
 Because of the superb spatial resolution
of {\it Chandra}, we were able to locate HoII X-1 with much better
accuracy than previous observation with {\it ROSAT} HRI (MLH01). A two-dimensional Gaussian fit of the Chandra image 
gives the position of HoII X-1 as RA: 8$^{h}$ 19$^{m}$ 29.0$^{s}$, DEC: +70$^{\circ}$ 42$^{\prime}$ 19$^{\prime\prime}$ (J2000). 
The ROSAT HRI position of RA: 8$^{h}$ 19$^{m}$ 29.7$^{s}$ and DEC: +70$^{\circ}$ 42$^{\prime}$ 18$^{\prime\prime}$
(J2000.0) agrees well within its uncertainties (Colbert \& Mushotzky \cite{Col99}).

Because there are no X-ray sources in the Chandra FOV which can be unambiguously 
identified with an optical/radio counterpart, the positional accuracy is 
limited by aspect uncertainties in the {\it Chandra} pointing. This is 
roughly estimated to be $\sim $1$^{\prime\prime}$, which we take as the radius 
of the positional error circle (Fig. \ref{FOV}). 

Three {\it XMM} observations of HoII X-1 were made with the EPIC PN/MOS detectors 
(see Table \ref{tab:xlog}). A total of 14.9 ks of good EPIC PN data were obtained after correcting for background flares. A combined events file was derived from the individual events files of the  different exposures. 
The {\it XMM} EPIC PN position of RA:  8$^{h}$ 19$^{m}$ 29.1$^{s}$ and DEC: +70$^{\circ}$ 42$^{\prime}$ 19$^{\prime\prime}$
(J2000.0) is in good agreement with the ACIS-S position.

 As already mentioned Tongue \& Westpfahl (\cite{Ton95}) have detected a radio peak at 6, 20 and 90 cm wavelengths at the position 8$^{h}$ 19$^{m}$ 28$^{s}$, +70$^{\circ}$ 42$^{\prime}$ 19$^{\prime\prime}$ (J2000.0). However, due to the low angular resolution of the radio data, it is not clear whether the radio emission $\sim3^{\prime\prime}$ offset from the X-ray position is coincident with the X-ray source.

\subsection{Spatial analysis}

In Fig. \ref{xray_cont} (left panel) we compare the radial profile of HoII~X-1 with the ACIS-S Point Spread Function (PSF) at 1 keV, where the peak of the source flux lies. 
The PSF was created using the pre-calculated profiles
available in the CIAO package (ver. 3.0) and normalized to the radial profile of the source at the first annulus outside 3$^{\prime\prime}$, 
due to heavy pile up in the central region. The 0.5-6 keV source profile was extracted avoiding possible serendipitous sources and the readout streak in the image.
 The background level was evaluated
from an annulus with innner$+$outer radius of 35$^{\prime\prime}+60^{\prime\prime}$, respectively. 
A hint of extended emission is detected between 4$^{\prime\prime}$-6$^{\prime\prime}$ from the source (also emphasized in Fig. \ref{xray_cont} (right panel), where a possible Westward extinction is evident. An upper limit on the contribution of the extended component to the total flux is estimated to be $\ll 15$\%.   
This value is a qualitative estimation of the extended component, since the pile up introduces an error $<20\%$. A longer exposure would be necessary to put a tighter constraint on the possible extended emission.

\subsection{Spectral Analysis} 

Here we present a first spectral analysis of the XMM-Newton public data. A more detailed analysis is provided in the
following paper by Dewangan et al. \cite{Dew04}. 

The EPIC-PN spectrum of HoII X-1 was extracted from a circular region of $\sim$150" in radius and the
background spectrum was obtained from a nearby source free region. For both spectra we use the combined events file from all three observations (see Table \ref{tab:xlog}). 
Firstly, we fitted a single power-law and neutral
absorption [fit (A)] to the PN spectrum in the energy range
from 0.3-10 keV. This gave a fair fit with a $\Gamma\sim 2.8$ power-law.
The fit parameters are shown in the first entry of Table \ref{tab:xfit}.
An obscured single black body, multi-colour disk black body (MCD), and thermal Bremsstrahlung models
gave an unacceptable fit ($\chi_{red}^{2}>2$).

\begin{figure}
\psfig{width=\hsize,file=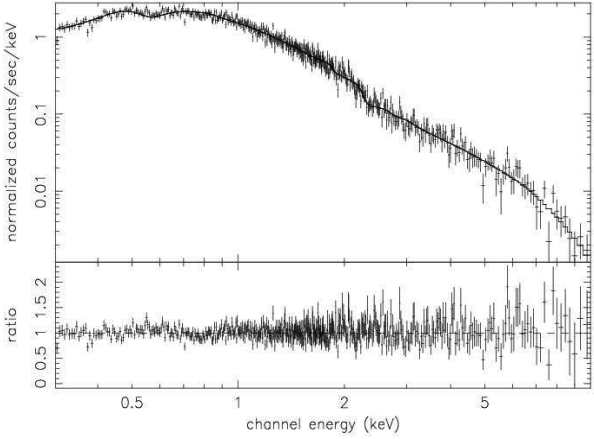}
\caption[]
 {The X-ray pulse-height spectrum of {\it XMM} EPIC-PN
 is shown with folded models for fit (B).
 The lower panel shows the data/model ratios. Pulse height spectra
 are re-binned for display and 1$\sigma$ error bars are attached to
 data points.  
 \label{fig:xfit}}
\end{figure}

\begin{figure}
\psfig{width=\hsize,file=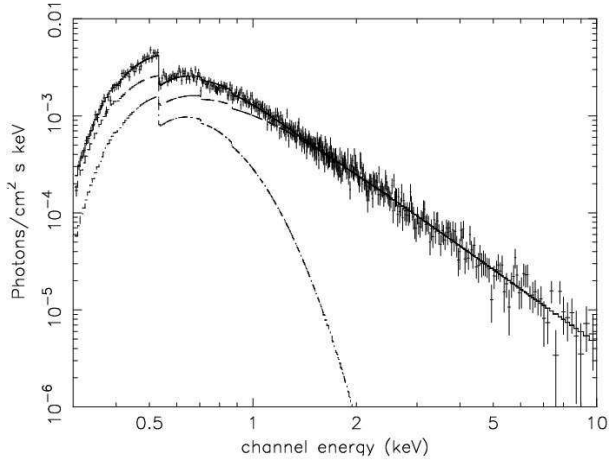}
\caption[]
 {Unfolded {\it XMM} EPIC-PN spectrum showing the best fit model (B). 
  The thermal black body component contributes about 18 \% to the total
  0.3-8.0 keV flux, whereas this component is clearly soft ($<2$ keV). 
 \label{fig:uxfit}}
\end{figure}

The fitted spectrum is  steeper than that found in our previous analysis of
the {\it ASCA} data, which gave $\Gamma\sim 1.9$.  In our previous 
analysis, however, we found a soft excess component in the joint spectral 
fit using {\it ASCA} and {\it ROSAT} data (MLH01). The best fit model to the {\it ASCA} and {\it ROSAT} data 
was a $kT\sim 0.3$ keV thermal plasma, or a $kT_{in}=0.17$ keV disk-blackbody, in addition to a $\Gamma\sim 1.9$
power-law. We have applied these models to the EPIC PN spectrum (see Fig. \ref{fig:xfit}) 
and found that the fit improved significantly. 
Thus, we consider our previous finding of at least two spectral components in HoII X-1 confirmed. The results of the power-law plus thermal fits
[fit (B)] are also shown in Table \ref{tab:xfit}. The thermal component is clearly soft (see Fig. \ref{fig:uxfit}) 
and contributes about 18 \% to the total 0.3-8.0 keV flux. 
A power-law fit to the soft component is ruled out ($\chi^{2}=566$ for 476 d.o.f.).  

An absorbed power-law and a multicolor disk black body (model C in Table \ref{tab:xfit}) gives a good fit as well. This means that the soft component can be as well described by a MCD model. The inner disk temperature is quite low ($kT\sim0.16$ keV), which indicate a cool accretion disk. 
Soft components with a cool accretion disks may indicate intermediate-mass black holes with a black hole mass of $M_{BH}\simeq10^{3}$M$_{\sun}$ (Miller et al. \cite{Mil03}).

\begin{table}[t]
\caption{Results of the EPIC PN spectral analysis}
\begin{tabular}{cl}
\hline\hline
Model & Parameters\\
\hline
{\bf PL$\cdot$Abs.} & $\Gamma=2.77^{+0.04}_{-0.04}$; $F_{\rm p12}=4.38^{+0.14}_{-0.13}$\\
   (A)              & $N_{\rm H20}=16.0^{+0.7}_{-0.6}$ \\
                    & ${\bf \chi^2/\nu=1.30}\,(621./478)$\\
           \\
{\bf PL+Thermal} (bbody)  &$\Gamma=2.50^{+0.09}_{-0.05}$; $F_{\rm p12}=3.38^{+0.36}_{-0.18}$; \\
{\bf  $\cdot$Abs.} & $kT=.14^{+0.01}_{-0.02}$;\\ 
   (B)             & $L_{39}/D^{2}_{10}=(2.84^{+1.64}_{-0.47})\cdot 10^{-5}$; \\
                   & $N_{\rm H20}=16.4^{+2.6}_{-1.1}$; \\
                   & ${\bf \chi^2/\nu=1.03}\,(492./476)$\\ 
           \\
{\bf PL+Thermal} (diskbb)  &$\Gamma=2.51^{+0.07}_{-0.07}$; $F_{\rm p12}=3.40^{+0.29}_{-0.26}$; \\
{\bf  $\cdot$Abs.} & $Tin=.16^{+0.02}_{-0.01}$;\\ 
   (C)             & $(R_{in}/D)^{2} cos \theta=407^{+516}_{-193}$ \\
                   & $N_{\rm H20}=18.7^{+2.3}_{-0.9}$; \\
                   & ${\bf \chi^2/\nu=1.04}\,(496./476)$\\ 
           \\
{\bf PL+Thermal} (mekal) &$\Gamma=2.65^{+.05}_{-.05}$; $F_{\rm p12}=3.97^{+0.23}_{-0.19}$; \\
{\bf  $\cdot$Abs.}  & $kT=.22^{+0.02}_{-0.02}$; $Z=1.0$;\\ 
   (D)             & $A_{\rm t}=(3.93^{+1.75}_{-0.77})\cdot 10^{-4}$; \\
                   & $N_{\rm H20}=16.2^{+1.3}_{-0.9}$; \\
                   & ${\bf \chi^2/\nu=1.06}\,(505./476)$\\ 
           \\
{\bf PL+Thermal} (mekal+ &$\Gamma=2.52^{+.06}_{-.08}$; $F_{\rm p12}=3.42^{+.22}_{-0.32}$; \\
bbody){\bf $\cdot$Abs.}  & $kT_{mekal}=.15^{+0.03}_{-0.04}$; $Z=1.0$;\\ 
   (E)                   & $A_{\rm t}=(4.6^{+10.3}_{-2.0})\cdot 10^{-4}$; \\
                         &  $kT_{bbody}=.15^{+0.01}_{-0.03}$;\\
                         & $L_{39}/D^{2}_{10}=(2.2^{+1.3}_{-0.6})\cdot 10^{-5}$; \\
                         & $N_{\rm H20}=16.9^{+1.3}_{-1.7}$; \\
                         & ${\bf \chi^2/\nu=0.98}\,(463./474)$\\
\\
\hline
\end{tabular}

\label{tab:xfit}

Fit parameters are shown with 90\% errors ($\Delta \chi^2=2.7$). 
The model and parameter definitions are:
Model Components-- {\bf PL}: Power-law with a photon index of $\Gamma$ 
and a 0.5-2 keV flux of $F_{\rm p12}$ in units of 10$^{-12}$ erg cm$^{-2}$ s$^{-1}$. 
{\bf Thermal -- thin plasma}: Thermal plasma using the XSPEC {\em mekal} model with 
a plasma temperature $kT$ [keV], a metal abundance $Z=1$ in solar units, 
and the normalization defined by XSPEC $A_{\rm th}$. 
{\bf Thermal -- multi color disk}: The multicolor disk (Mitsuda et al. \cite{Mit84}) distributed
as the XSPEC model (diskbb) with an inner disk temperature $kT_{in}$ and with an normalization defined as
$(R_{in}/D) cos \theta$, where $R_{in}$ is the inner disk radius and $D$ is the distance to the source in 10 $kpc$ units, and $\theta$ is the viewing angle of the disk axis. 	
{\bf Thermal -- thick plasma}: Thermal thick plasma using the XSPEC {\em bbody}
 model with a plasma temperature $kT$ [keV], with a normalization defined as L$_{39}$/D$^{2}_{10}$,
where L$_{39}$ is the source luminosity in units of 10$^{39}$ erg s$^{-1}$ and D$_{10}$ is the distance to the source in units of 10 kpc.
{\bf Abs:} Absorption by neutral gas using the XSPEC model {\em wabs} 
(Morrison R. McCammon D. 1983, ApJ 270 190) with hydrogen column density 
$N_{\rm H20}$ $[10^{20}{\rm cm}^{-2}]$.
\end{table}

The excess of cold absorption ($N_{\rm H}\sim 1.6\times 10^{21}$ cm$^{-2}$) over
the galactic value ($N_{\rm H}\sim 3.4\times 10^{20}$ cm$^{-2}$) is confirmed by all fits (A to D).
The intrinsic cold absorption is in good agreement with recent VLA observations of Holmberg II.
Bureau \& Carignan (\cite{Bu02}) showed that the compact H II region \#70 is located inside one 
of the regions with the largest intrinsic hydrogen column density (1.9$\times 10^{21}$ cm$^{-2}$).

A power-law and a thermal thin plasma model using a solar metal abundance [fit (D)] also gives a resonable fit.
The fraction of the 0.3-8.0 keV thermal thin flux (mekal) to the total intrinsic flux is about 9 \%.
The thick plasma thermal emission (bbody) is assumed to be originating from the accretion disk around the black hole.
Therefore, the thick plasma thermal emission is spatially not resolved. The spatial analysis of the {\it Chandra}
ACIS-S image suggests a possible extended component with a fraction of less than 15 \% to the total 0.3-8.0 keV flux, which would be in agreement with the flux fraction resulting from the thermal thin plasma. This would favor fit (D). 

A power-law model and a combination of a thick and a thin thermal plasma model (E) gives
only a marginal improvement of the fit.
The intrinsic flux in the 0.3-8.0 keV band is 9.2$\times10^{-12}$ erg s$^{-1}$ cm$^{-2}$ corresponding to
a luminosity of 1.1 $\times$ 10$^{40}$ erg s$^{-1}$. The contribution of the power-law component, the thermal black body component, and the thermal thin plasma component to the total 0.3-8.0 keV flux are 77 \%, 16 \% and 7\%, respectively. However, a longer EPIC-PN exposure is needed to disentangle a possible extended thermal thin plasma from the point-like power-law and thermal thick plasma components.

The spectral analysis of the combined XMM data has confirmed two spectral components in HoII X-1.
The hard component is best described by a powerlaw ($\Gamma\sim2.6$), and the soft component is fitted by  thermal models with relatively low temperatures ($kT\sim0.14-0.22$ keV). The intrinsic X-ray luminosity (assuming isotrophic emission) is about 10$^{40}$ erg s$^{-1}$ in the 0.3-8.0 keV band.

\section{Discussion}

Before we discuss the nature of the ultraluminous X-ray source HoII X-1 we summarize our results:

\begin{enumerate}
\item{An extended He II $\lambda4686$ region (21$\times47$ pc) inside an H II region is positionally coincident with the HoII X-1 source. The existence of such an extended He~II emission region is very rare in H II regions. The He II luminosity agrees well with those published by Pakull \& Mirioni (\cite{Pak01}) and Kaaret et al. (\cite{Kaa04}).}
\item{Neither the emission line fluxes nor the line flux ratios of the lines other than He II $\lambda4861$ are correlated with HoII X-1. The emission line flux ratios agree well with those of normal H II regions. This result excludes a SNR nature of the radio source inside the Holmberg II region \#70, which is probably not coincident with the X-ray source. }
\item{We find a blue extended counterpart for the ULX with a size of 11$\times14$ pc, which is most probably a young stellar complex or a cluster. We have derived an X-ray to optical luminosity ratio of L$_{X}/$L$_{B}\ge170$ for the blue stellar complex. The recently published {\it HST} ACS images of HoII X-1 
      (Karret et al. \cite{Kaa04}) confirm this, however they suggest a ratio of L$_{X}/$L$_{B}=300-400$.}
\item{The increased FWHM of $\sim50$ km s$^{-1}$ in a scale of $\sim20-30$ pc at the position of HoII X-1 is most likely not directly gravitational related  to the rotation around a black hole, as the resulting black hole mass would be $\sim10^{7}$M$_{\sun}$. 
Further, there is a radial velocity variations inside the He II region of $\pm50$ km s$^{-1}$ on spatial scales of $\sim2^{\prime\prime}$ ($\pm30$ pc in projection). The complex velocity field at the location of the ULX suggests that the putative black hole ionizes the surrounding HII gas and perturbs it dynamically (via jets or the accretion disk wind).}
\item{The spatial analysis of the {\it Chandra} ACIS-S image gives marginal indications for an extended emission component with an extended flux fraction well below 15 \%. This could be consistent with the thin thermal plasma component in the {\it XMM} spectrum. }
\item{The {\it XMM} EPIC-PN spectrum is well fitted  with an absorbed power-law ($\Gamma\sim2.6$) for the hard
      component and a thermal emission for the soft component, which can be due to a thick plasma (bbody), a thin plasma (mekal), or a multicolor black body disk with relatively low inner disk temperature ($kT_{in}\sim0.16$      keV). The intrinsic absorption 
      of N$_{H}\sim1.6 \times 10^{21}$ cm$^{-2}$ is greater than the galactic absorption. } 
\end{enumerate}

There are three models for ULX extensively discussed  in the literature:

i) ULX may be black holes of ''normal'' stellar masses ($\sim 10$M$_{\sun}$)
in binaries, which accrete gas in a supercritical regime. They are
SS\,433-like objects or microquasars in their transient activity states,
whose hard radiation can be collimated (and beamed) along jets and
accretion disks axes (Fabrika \& Mescheryakov, \cite{Fa01}, King et al., \cite{Ki01}).

ii) ULX may be black holes with few tens of solar mass ($\sim 10$M$_{\sun}$)
and that their X-ray emissions is from the disk shining at super-Eddington luminosities
(Begelman \cite{Beg02}). The slim disk model (Abramowicz et al. \cite{Abra88}, Ebisawa et al. \cite{Ebi03}) 
allows to explain the observed super-Eddington luminosity, hard X-ray spectra and spectral
variations. 

iii) The ULX may be intermediate mass black holes (IMBHs), $\sim 10^3$M$_{\sun}$
(Colbert \& Mushotzky \cite{Col99}), which were formed from the very first stars
(Madau \& Rees \cite{Mad01}) or in globular clusters (Miller \& Hamilton \cite{Mil02}).
The IMBH could accrete gas from a close companion or even from the interstellar medium and become bright X--ray sources
when they are in dense gas environments. 
Spectra of some ULX (Miller et al. \cite{Mil03}) support the idea that they are IMBHs.

Substantial extended He II~$\lambda4886$ emission has been confirmed inside the compact H II region \#70 at the position of HoII X-1. Such strong He II~$\lambda4886$ emission is produced by reprocessing of the X-ray source located inside the nebula.   
 Pakull \& Mirioni (\cite{Pak01}) have concluded that the optical radiation is consistent with a quasi-isotrophic X-ray source and no significant beaming is
required. In this case  the source is really so bright in X--rays (L$_{X}\sim 10^{40}$\,erg/s), which supports an IMBH interpretation ($iii$). 
The extended blue counterpart of HoII X-1 suggests that the IMBH is
located in a young stellar complex or cluster.

The IMBH may disturb the velocity field (via the accretion disk wind or jets) of the H II region at its position, where we have found an increase of the velocity dispersion, and a radial velocity variation accross the He II~$\lambda4686$ region. The amplitude of the radial velocity gradient accross the He II~$\lambda4686$ region is $\pm50$ km s$^{-1}$ on spatial 
scales of $\pm2^{\prime\prime}$. This means that the ULX could not have only ionized the surrounding gas, but also has interacted with it dynamically.

The X-ray properties of HoII X-1 seem to agree with the IMBH interpretation ($iii$).  The soft component of the X-ray spectrum of HoII X-1 can be described by a MCD model with a inner disk temperature of $kT\sim0.16$ keV. Such a cool accretion disk indicates intermediate-mass black holes 
with a black hole mass of $M_{BH}\simeq10^{3}$M$_{\sun}$ (Miller et al. \cite{Mil03}). A further indication for a IMBH is the long-term variability of HoII X-1 (see MLH01 and Dewangan et al. \cite{Dew04}).

The optical/X-ray properties of HoII X-1 seem to favour  the IMBH nature of the object.  High angular resolution integral field observation with HST or ground based near-infrared integral field spectroscopy using adaptive optics  at an 8-m class telescope are needed to prove the IMBH nature of HoII X-1, and would allow to put tighter constrains on the velocity structure of the gas and the stars inside the cluster around HoII X-1, and could enable us to measure the mass of the IMBH.

\begin{acknowledgements}
T. Becker and M.M. Roth were Visiting Astronomers, German-Spanish Astronomical Centre, Calar Alto, operated by the Max-Planck-Institute for Astronomy, Heidelberg, jointly with the Spanish National Commission for Astronomy.

The authors thank the Canadian Astronomy Data Center, which is operated by the
Dominion Astrophysical Observatory for the National Resarch Council
of Canada's Herzberg Institute of Astrophysics.
This research has made use of the USNOFS Image and Catalogue Archive
operated by the United States Naval Observatory, Flagstaff Station
(http://www.nofs.navy.mil/data/fchpix/).
The work has been supported by Russian RFBR grants N\,03--02--16341, 04--02--16349, and INTAS grant YSF 2002-281.

We thank L. Gallo for his valuable comments and suggestions.
Further we want to thank V. Goranskii for help with the photometry of the optical counterpart.
\end{acknowledgements}

\end{document}